\documentclass[oldversion]{aa}  % non-structured abstract

\usepackage{graphicx}
\usepackage{txfonts}
\usepackage{aalongtable}
\usepackage{float}
\usepackage[latin1]{inputenc}
\usepackage{supertabular}
\usepackage{natbib}
\bibpunct{(}{)}{;}{a}{}{,}

\begin{document}

\title{The Hamburg/ESO R-process Enhanced Star survey (HERES) \thanks{Based on
    observations collected at the European Southern Observatory, Paranal,
    Chile (Proposal Number 170.D-0010).}  }

\subtitle{XI. The highly $r$-process-enhanced star CS~29497-004}

\author{
  V. Hill\inst{1} \and
  N. Christlieb\inst{2} \and
  T.C. Beers\inst{3} \and
  P.S. Barklem\inst{4} \and
  K.-L. Kratz\inst{5,6} \and
  B. Nordström\inst{7,8}
  B. Pfeiffer\inst{6}
  K. Farouqi\inst{5,6}
}

%1) I cannot speak for Bernd Pfeiffer. He has contributed a lot to the "old" waiting-point
%   approach.
%   If remaining co-author,take my below Mainz Univ. affiliation, and as email adress 
%   Dr,_Bernd-Pfeiffer@t-online.de.
%
%2) My present TWO official affiliations are:
%   Max-Planck-Institut fr Chemie (Otto-Hahn-Institut), D-55128 Mainz;   k-l.kratz@mpic.de, 
%   and
%   Fachbereich Chemie, Pharmazie & Geowissenschaften, Universitt Mainz,  D-55128 Mainz; 
%   klk@uni-mainz.de.
%
%3) My suggestion for the present / final paper version is to include Khalil Farouqi as
%   author -- for his "new" HEW work.
%   Part of the time of this work, he was a member of Norbert's group in Heidelberg. Ask
%   Norbert for the HD-affiliation.
%   With respect to Mainz, tke BOTH of my affiliations; and as email adress: farouqi@live.de.

\offprints{V. Hill,\email{vanessa.hill@oca.eu}}

\institute{
Laboratoire Lagrange, Universit\'e de Nice Sophia-Antipolis, Observatoire de la C\^ote d'Azur, CNRS, Bd de l'Observatoire, CS 34229, 06304 Nice cedex 4, France; 
     \email{vanessa.hill@oca.eu} % 1
\and Zentrum für Astronomie der Universität Heidelberg, Landessternwarte,
     Königstuhl 12, 69117 Heidelberg, Germany; 
     \email{N.Christlieb@lsw.uni-heidelberg.de} %2
\and Department of Physics and JINA Center for the Evolution of the Elements,\\
     University of Notre Dame, Notre Dame, IN 46556, USA;
     \email{tbeers@nd.edu} %3
\and Theoretical Astrophysics, Department of Physics and Astronomy, Uppsala University, Box 516, SE-751 20 Uppsala, Sweden; \email{Paul.Barklem@physics.uu.se} %4
\and Max-Planck-Institut für Chemie (Otto-Hahn-Institut), D-55128 Mainz; %5
\and Fachbereich Chemie, Pharmazie \& Geowissenschaften, Universität Mainz,  D-55128 Mainz;  \email{klk@uni-mainz.de, Dr.$\_$Bernd-Pfeiffer@t-online.de, farouqi@live.de} %6
\and Dark Cosmology Centre, The Niels Bohr Institute, Copenhagen University, Juliane Maries Vej 30, DK-2100, Copenhagen, Denmark; \email{birgitta@nbi.ku.dk}%7
\and Stellar Astrophysics Centre, Department of Physics and Astronomy, Aarhus University, Ny Munkegade 120, DK-8000, Aarhus C, Denmark%8
}

\date{Received  / Accepted }

\abstract{

We report an abundance analysis for the highly $r$-process-enhanced
($r$-II) star \object{CS~29497-004}, a very metal-poor giant with
$T_{\mathrm{eff}}$ = 5013~K and [Fe/H] $= -2.85$, whose nature was
initially discovered in the course of the HERES project. Our analysis is
based on high signal-to-noise, high-resolution ($R \sim 75000$) VLT/UVES
spectra and MARCS model atmospheres under the assumption of local
thermodynamic equilibrium, and obtains abundance measurements for a
total of 46 elements, 31 of which are neutron-capture elements. As is the
case for the other $\sim 25$ $r$-II stars currently known, the
heavy-element abundance pattern of \object{CS~29497-004} well-matches a
scaled Solar System second peak $r$-process-element abundance pattern. We confirm
our previous detection of Th, and demonstrate that this star does not
exhibit an ``actinide boost''. Uranium is also detected
($\log\epsilon(\mathrm{U}) =-2.20\pm0.30$), albeit with a large
measurement error that hampers its use as a precision cosmo-chronometer.
Combining the various elemental chronometer pairs that are available for
this star, we derive a mean age of $12.2\pm 3.7$\,Gyr using the
theoretical production ratios from waiting-point approximation models
(Kratz et al. 2007). %while the U/Th ratio leads to an age of $16.2\pm 3.7$\,Gyr
We further explore  the high-entropy wind model (Farouqi et al. 2010) production 
ratios arising from different neutron richness of the ejecta ($Y_e$), and derive an age of 
 $13.7 \pm 4.4$\,Gyr for a best-fitting $Y_e=0.447$. The U/Th nuclei-chronometer is confirmed 
 to be the most resilient to theoretical production ratios and yields an age of $16.5 \pm 6.6$\,Gyr. 
Lead (Pb) is also tentatively detected in \object{CS~29497-004}, at a level
compatible with a scaled Solar $r-$process, or with the theoretical expectations for a pure $r$-process in this star. 

}

\titlerunning{Detailed abundance analysis of CS~29497-004}
\authorrunning{Hill et al.}

\maketitle

%----------------------------------------------------------------------------%
\section{Introduction}\label{Sect:Intro}

Even though the physics of the rapid neutron-capture process
($r$-process) has been explored since the seminal works of
\citet{B2FH:1957} and \citet{Cameron:1957},
the astrophysical site(s) and stellar progenitor(s) that could give rise
to the extreme conditions needed for its onset remain elusive. The
discovery of the first highly $r$-process-element enhanced metal-poor
star, \object{CS~22892-052} \citep{Snedenetal:1994}, and the remarkable
agreement of its $r$-process-element abundance pattern with the
(inferred) Solar System pattern, prompted vigorous exploration of models
in which individual core-collapse supernovae are possible sites of the
astrophysical $r$-process (see \citealt{Snedenetal:2008} for a review).
Subsequent searches for such stars, and their detailed spectroscopic
follow-up, have revealed that the scaled-Solar heavy-element abundance
pattern for very ([VMP: Fe/H] $< -2.0$ and extremely (EMP: [Fe/H] $<
-3.0$) metal-poor stars with [Eu/Fe] $> +1.0$ and [Ba/Eu] $< 0.0$
($r$-II stars: \citealt{Beers/Christlieb:2005}) is extremely robust, and
provides what has come to be known as the ``universality constraint" on
the origin of the $r$-process. Recent theoretical attention has focused
on the possibility that merging neutron star (and/or neutron star /
black hole) binaries may play a central role in the early production of
the $r$-process elements that satisfies the universality constraint
\citep[e.g., ][and references therein]{Ishimaruetal:2015}, in particular if they
took place in astrophysical environments in which star formation and
mixing (dilution) was inefficient. This latter suggestion has been given
recent observational support from the discovery of a substantial
population of $r$-II stars in the ultra-faint dwarf galaxy Reticulum II
\citep{Jietal:2016,Roedereretal:2016}, a galaxy which may be similar to the
low-mass mini-halos that contributed to the early chemical enrichment of
the halo of the Milky Way.
  
Besides probing specific alternative progenitors and environments by
studies of their observed $r$-process abundance patterns, measurements
of the abundances of long-lived radioactive isotopes for $r$-II stars,
such as thorium ($^{232}$Th; {half-life $\sim$ 14\,Gyr) and uranium
($^{238}$U; half-life $\sim$ 4.5\,Gyr), can also place constraints on the
age of the Universe \citep[e.g., ][]{Cayreletal:2001, Hilletal:2002}. It
must be recognized, however, that $r$-II stars are exceedingly rare --
comprising no more than 3-5\% of {\it all} VMP and EMP stars. Only two $r$-II
stars with unambiguously detected U have been reported in the
literature, \object{CS~31082-001} \citep{Hilletal:2002} and
\object{HE~1523-0901} \citep{Frebeletal:2007}. 

While the first few $r$-II stars were identified from high-resolution
spectroscopic follow-up of VMP and EMP stars selected from the HK survey
\citep{BPSI, BPSII}, the Hamburg/ESO R-process Enhanced Star
(HERES) survey was the first dedicated effort to substantially increase
the numbers of known $r$-II (and $r$-I, $+0.3 \le$ [Eu/Fe] $\le +1.0$
and [Ba/Eu]$ < 0.0$; \citealt{Beers/Christlieb:2005}) stars in the halo
of the Milky Way. Its motivations have been described in previous papers
of this series (\citealt{HERESpaperI}, hereafter Paper~I;
\citealt{HERESpaperII}, hereafter Paper~II). In Paper~I, we also
reported on the discovery of the highly $r$-II star
\object{CS~29497-004}, and the detection of the \ion{Th}{II} line at
$\lambda = 4019.129$\,{\AA} in its HERES ``snapshot'' spectrum. Herein,
we demonstrate that it is now the third metal-poor star known with
clearly detected uranium, based on a detailed abundance analysis of
high-quality VLT/UVES spectra using MARCS model atmospheres. 

For the convenience of the reader, we have summarized the coordinates
and photometry of \object{CS~29497-004} in Table
\ref{Tab:CoordsPhotometry}. 

This paper is outlined as follows. The observations and data processing
are described in Sect. \ref{Sect:Observations}, the determination of the
stellar parameters in Sect. \ref{Sect:StellarParameters}, and the
abundance analysis in Sect. \ref{Sect:AbundanceAnalysis}. The results
are summarized in Sect. \ref{Sect:Results} and discussed in Sect.
\ref{Sect:DiscussionConclusions}.

\begin{table}[htbp]
 \centering
 \caption{\label{Tab:CoordsPhotometry} Coordinates and photometry of 
          \object{CS~29497-004}. References: 1 -- Paper~I; 
          2 -- \citet{Beersetal:2007}; 
          3 -- \citet{Skrutskieetal:2006}.}
 \label{tab:CoordsPhotometry}
  \begin{tabular}{lll}\hline\hline
   Parameter    & Value              & Ref.\\\hline
   R.A. (2000.0) & $00:28:06.9$       & 1   \\
   DEC  (2000.0) & $-26:03:04$        & 1   \\
   $V$ [mag]    & $14.034 \pm 0.005$ & 2   \\
   $B-V$ [mag]  & $ 0.705 \pm 0.010$ & 2   \\
   $V-R$ [mag]  & $ 0.451 \pm 0.006$ & 2   \\
   $V-I$ [mag]  & $ 0.918 \pm 0.007$ & 2   \\
   $J$   [mag]  & $12.491 \pm 0.021$ & 2,3 \\
   $H$   [mag]  & $12.038 \pm 0.024$ & 2,3 \\
   $K$   [mag]  & $11.965 \pm 0.025$ & 2,3 \\\hline
  \end{tabular}
\end{table}

\section{Observations, data reduction, and radial-velocity variation}\label{Sect:Observations}

High-resolution spectra of \object{CS~29497-004} were obtained with UVES at
VLT-UT2 during October--December 2002 and July--August 2003. A total of 22 exposures were
obtained with the BLUE346 setting, and 6 with the BLUE437 setting. In
both cases Image Slicer \#2 was used, yielding a nominal resolving power of
$R=\lambda/\Delta\lambda=75,000$. The total useful wavelength range covered by
these two settings is 3050--4980\,{\AA} in the rest frame of the star. The
observations are summarized in Table \ref{Tab:Observations}.

\begin{table}[htbp]
 \centering
 \caption{\label{Tab:Observations} UVES observations of \object{CS~29497-004}.
          $\lambda$ refers to the combined useful wavelength range (in the 
          rest frame of the star) of all $N$ individual exposures, and $t$ is
          the total integration time.}
  \begin{tabular}{llrr}\hline\hline
   Setting & $\lambda$ [{\AA}] & \multicolumn{1}{l}{$N$} & 
   \multicolumn{1}{l}{$t$ [h]}\\\hline
   BLUE346 & 3050-3865 & $22$ & $21.2$\\
   BLUE437 & 3760-4980 & $ 6$ & $ 4.9$\\\hline
  \end{tabular}
\end{table}

The individual pipeline-reduced spectra were first corrected for
geocentric radial-velocity shifts, determined by measuring the positions
of $\sim 10$ clean absorption lines throughout the covered spectral
range. Then, the spectra at each setting were co-added using an iterative
procedure in which pixels in the individual spectra affected by cosmic
ray hits not fully removed during the data reduction, or by by CCD
defects or other artifacts, were identified. These pixels were flagged
and ignored in the final iteration of the co-addition procedure. The
co-addition takes into account the $S/N$ of the individual spectra (the
noise estimate being provided by the UVES pipeline), by computing for
each final spectral bin a weighted average of the corresponding bins
in the individual spectra.

The resulting co-added spectrum of the BLUE346 setting has a peak $S/N$
of $\sim 150$ per 0.0125\,{\AA} pixel in the reddest part of the
spectrum, roughly linearily decreasing to $S/N=50$ at 3300\,{\AA} and
$S/N=10$ at 3100\,{\AA}. The co-added BLUE437 spectrum has a maximum
$S/N$ per 0.0152\,{\AA} pixel of 120 at 4980\,{\AA}, decreasing to $\sim
60$ at 3760\,{\AA}.

In addition, all 28 spectra were co-added in the wavelength range
covered by both settings, 3760--3865\,{\AA}. The $S/N$ did not increase
significantly with respect to the 22 co-added BLUE346 spectra, but the
additional co-added spectrum has been used for verifying the detection
of critical lines such as the U~II 3859.57\,{\AA} line.

Together with the UVES ``snapshot'' spectrum taken on 1 November 2001
(see Paper~I), the UVES spectra cover a period of about 3 years. During
this period of time, no significant radial-velocity variations of
\object{CS~29497-004} were detected (see Fig. \ref{Fig:vrad_CS29497}).
\citet{Hansenetal:2015} reported on 12 independent high-resolution
observations of this star, extending over a temporal window of seven years, and
obtained a mean velocity of 105.0 $\pm 0.37$ km/s, which agrees well with our
own measurements.  When combined with our present measurements and
those from Paper~I, the temporal window over which this star has not exhibited
significant radial-velocity variations expands to  $\sim13$ years.
These data have been examined in detail, and no suitable orbit could be
identified (T.T. Hansen, private communication). It thus appears that
\object{CS~29497-004} is a single star, consistent with the great
majority of $r$-II (and  $r$-I) stars in the \citet{Hansenetal:2015}
study. These authors report that the binary fraction of
$r$-process-element enhanced stars is 18 $\pm 11$\%\footnote{Note that
the published error bar on this fraction was incorrectly reported as
6\%.}, commensurate with that found for other ``normal" halo giants.

\begin{figure}[htbp]
  \centering
  \includegraphics[width=\hsize,bb=72 568 399 729, clip= ]{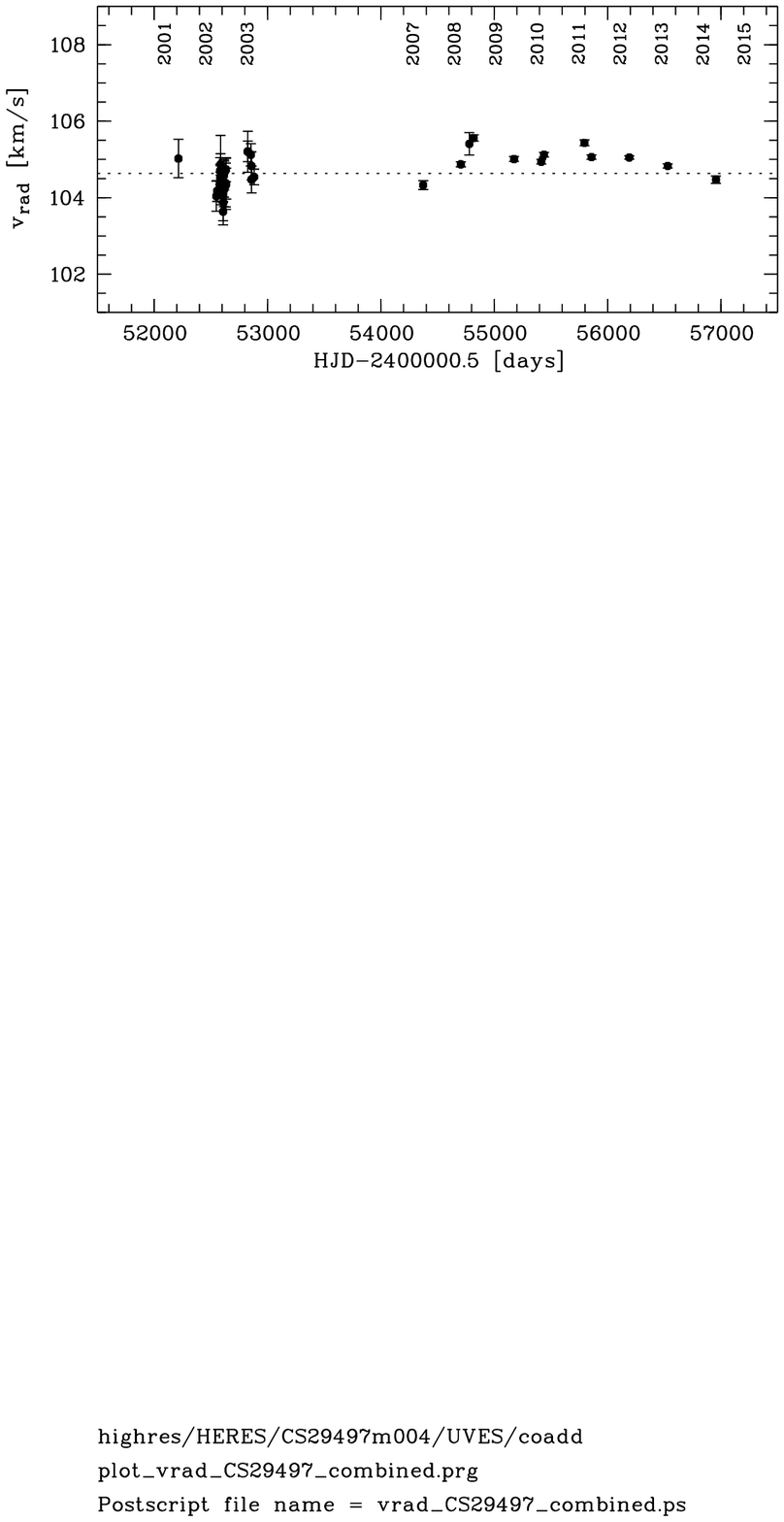} 
  \caption{\label{Fig:vrad_CS29497} Barycentric radial velocities of
  CS~29497-004 from the present paper (observations in 2002-2003), those
  from Paper~I (observations in 2001), and from \citet{Hansenetal:2015}
  (observations in 2007 to 2014). See Table~\ref{tab:vrad} for a
  complete list of these data. No significant radial velocity variations
  were detected on a timescale of $\sim 13$ years.}
\end{figure}

\section{Stellar parameters}\label{Sect:StellarParameters}

The stellar parameters of \object{CS~29497-004} adopted throughout this
paper are reported in Table~\ref{tab:StellarParameters}; they were
obtained in the following manner. The effective temperature,
$T_{\mathrm{eff}}$, that was derived in Paper II, based on a combination
of $BVRI$ and $JHK$ broadband colours and the colour-temperature
calibrations of \cite{Alonsoetal:1999a}, was verified to successfully
fulfill the excitation equilibrium for a sample of 77 \ion{Fe}{i} lines,
and subsequently adopted. Surface gravity, $\log g$, was deduced from
the ionisation equilibrium of \ion{Fe}{i} and \ion{Fe}{ii} lines. The
microturbulence velocity, $v_{\mathrm{micr}}$, was deduced by requiring
that \ion{Fe}{i} lines of all strengths produced the same Fe abundance.  
 Fig.~\ref{Fig:Feplot} illustrates the adequacy of the stellar parameters 
to reproduce excitation and ionisation balance as well as a flat behaviour 
of the abundance with increasing line strength.

\begin{figure}[htbp]
  \centering
  \includegraphics[width=\hsize, clip= ]{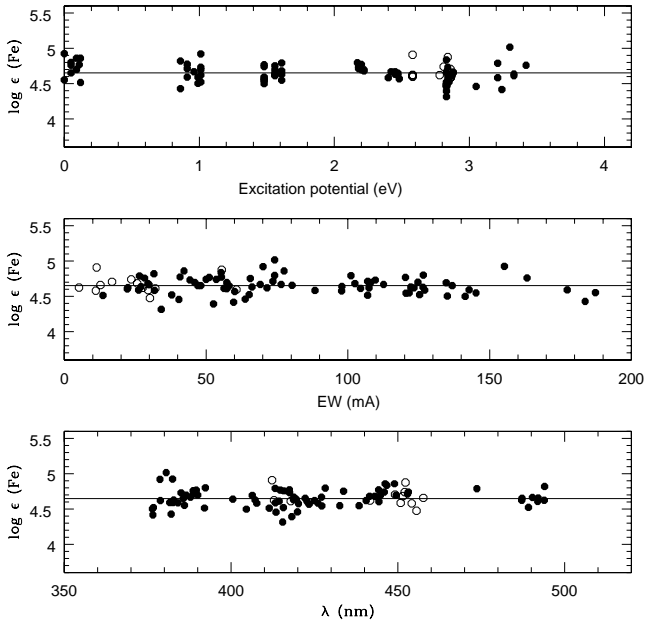} 
  \caption{\label{Fig:Feplot} Iron abundances (\ion{Fe}{i} filled circles, \ion{Fe}{ii} open circles) as a function of excitation potential, equivalent width, and line wavelengths, showing the adequacy of the stellar parameters to reproduce excitation and ionisation balance as well as a flat behaviour of the abundance with increasing line strength.}
\end{figure}

\begin{table}[htbp]
 \centering
 \caption{Stellar parameters of \object{CS~29497-004}.} 
 \label{tab:StellarParameters}
 \begin{tabular}{lllll}\hline\hline
                              & Paper~I        & Paper~II       & This work\\\hline
   $T_{\mathrm{eff}}$ [K]       & $5090\pm100$   & $5013\pm100$   & $5013\pm100$ \\
   $\log g$ (cgs)             & $2.4\pm0.4$    & $2.23\pm0.24$  & $2.05\pm0.1$ \\
   $\mbox{[Fe/H]}$            & $-2.64\pm0.12$ & $-2.81\pm0.13$ & $-2.85\pm0.11$ \\
   $v_{\mathrm{micr}}$ [km/s]   & $1.6\pm0.2$    & $1.62\pm0.13 $ & 1.60$\pm0.1$ \\\hline
  \end{tabular}
\end{table}

\section{Abundance analysis}\label{Sect:AbundanceAnalysis}

\subsection{Model atmospheres}\label{Sect:ModelAtmospheres and codes}

The model atmospheres used in this analysis are the same as those used
in the analysis of the $r$-II star \object{CS~31082-001}
\citep{Cayreletal:2001,Hilletal:2002} and the VMP/EMP
giants in the First Stars project \citep{Cayreletal:2004}. The models
are interpolated in a grid of OSMARCS models computed between 2000 and
2002 by B. Plez, using the latest version of the MARCS code. The MARCS
code was originally developed by \citet{Gustafssonetal:1975}, and updated
and improved over the years by \citet{Plez:1992},
\citet{Edvardssonetal:1993}, \citet{Asplundetal:1997}, and
\citet{Gustafssonetal:2008}.

The abundance measurements themselves were carried out using the suite
of programs {\em Turbospectrum} \citep{Alvarez/Plez:1998}, either
through an equivalent-width analysis, or direct comparison (and
$\chi^2$ minimisation) between the observed and synthetic spectra. The
abundances for elements from Mg through Zn were computed using
equivalent widths, while the abundances of C, N, O, and of most of the
neutron-capture elements, are measured by comparing directly the observed
spectrum with synthetic spectra.

\subsection{Line data}\label{Sect:LineData}

The linelist adopted to determine abundances in this paper is the same
as the one used in \citet{Hilletal:2002} and \citet{Cayreletal:2004},
with the exception of a number of neutron-capture elements for which improved 
atomic data has appeared in recent years, thanks to the efforts
of atomic physicists \citep[e.g.,][]{Lawleretal:2001a,Lawleretal:2001b,
Lawleretal:2004,Lawleretal:2006, Lawleretal:2007,Nilssonetal:2002a,
Nilssonetal:2002b}. The neutron-capture element linelist is given in the
Appendix (Table~\ref{Tab:LinelistHeavy}), together with the references
for transition probabilities and hyperfine structure coefficients, where
appropriate.  
%<-- ANY NEW N-CAPTURE ELEMENT DATA AVAILABLE NOW ?
% only Lawler 2008 missing for ErII.... I may update while the paper is in process of referee.

The atomic linelist used for the spectrum synthesis was assembled using
the VALD database \citep{Kupkaetal:1999,Kupkaetal:2000}, and oscillator
strength and/or hyperfine structures were updated for specific lines.
The same molecular linelists as in \citet{Hilletal:2002},
\citet{Spiteetal:2005}, and \citet{Spiteetal:2006} were included in the
syntheses for $^{12}$CH, $^{13}$CH, $^{12}$C$^{14}$N, $^{13}$C$^{14}$N,
$^{14}$NH, $^{15}$NH, and $^{16}$OH.

\subsection{Equivalent-width measurements}\label{Sect:EqwAnalysis}

Equivalent-width measurements were performed independently by V.H. and
N.C.; Fig.~\ref{Fig:EqwTest} shows a comparison of these two sets of
measurements. They agree very well for lines weaker than about 100\,
m{\AA}. Due to the fact that different relative weights of the line core
and wings are implicitly assigned by the fitting procedures used, the
equivalent widths of strong lines measured by N.C. are systematically
higher than those measured by V.H. In any case, for \emph{both}
procedures the equivalent widths of strong lines may be underestimated,
because Gaussian profiles were used. Ideally, Voigt profiles should be
used for the fits to strong lines, since they exhibit damping profiles.

\begin{figure}[htbp]
  \centering
  \includegraphics[width=\hsize,bb=70 285 314 598,clip=]{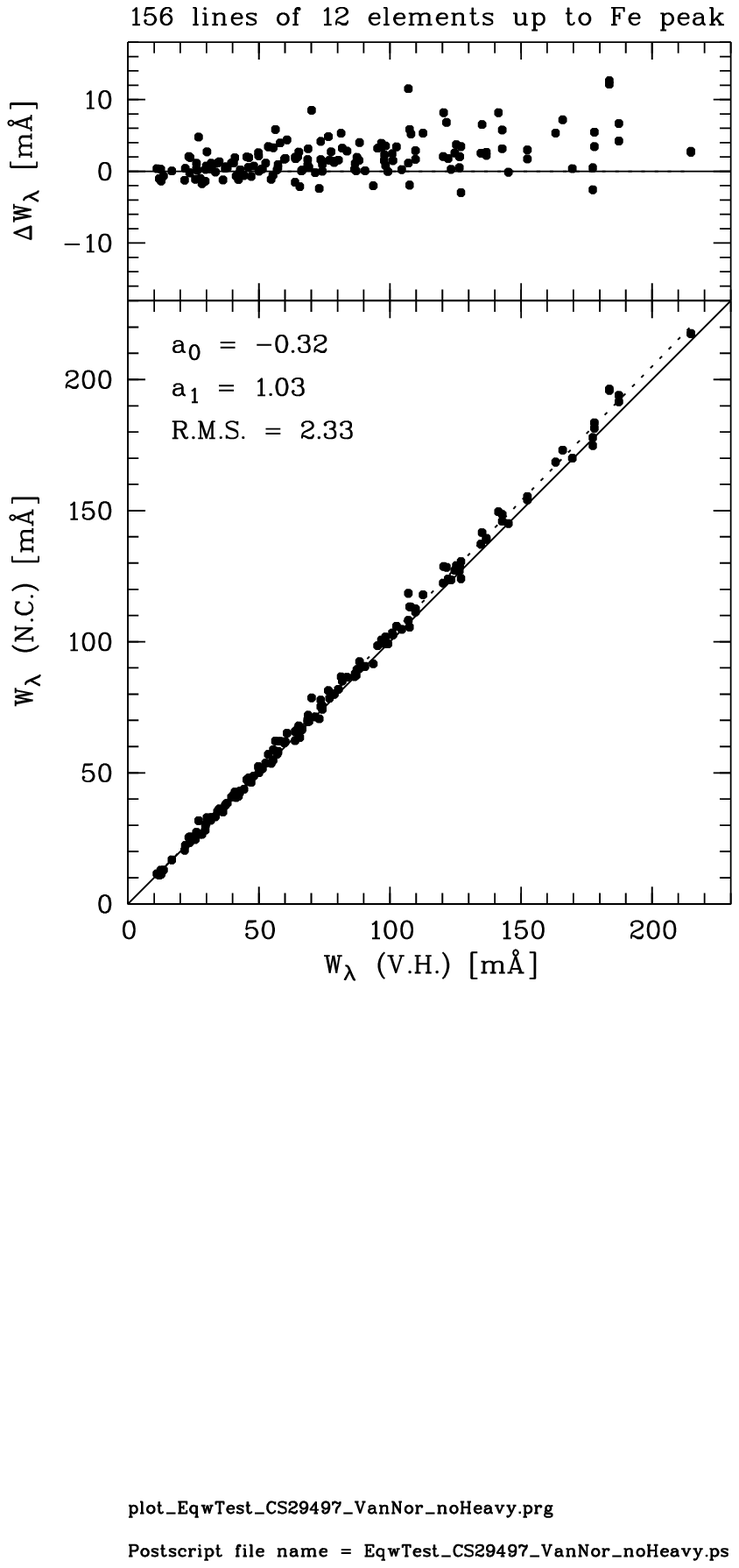}
  \caption{\label{Fig:EqwTest} Equivalent widths for 156 lines of 12
elements, up to the Fe peak.  The initials on each axis indicate the
authors who (independently) performed these measurements.  The solid line
is the one-to-one line, while the dashed line indicates a linear fit to 
the data.}
\end{figure}

In order to be consistent with the analysis of \object{CS~31082-001}
\citep{Cayreletal:2001,Hilletal:2002}, we adopted the measurements of V.H. 
for the equivalent-width analysis. 

\subsection{C, N, and O abundances}\label{Sect:MolSynthesis}

The elements carbon, nitrogen and oxygen are detected via molecular bands of CH, NH, and
OH.  Molecular equilibrium is solved including all relevant molecules for the atmosphere 
of cool stars, including CO which plays a major role in the C,N,O molecular equilibrium.

The carbon abundance is derived primarily from CH lines in the region
4290-4310\,{\AA} (the CH A-X 0-0 bandhead, often referred to as the
$G$-band), which is almost free from intervening atomic lines. We derive
an abundance $\rm\log\,\epsilon\,(C)= 5.94\,\pm\,0.5$ (see Fig.
\ref{Fig:Gband}). With this same abundance, a good fit is obtained in
the more crowded regions around 3900\,{\AA} and 3150\,{\AA}, where the
B-X and C-X bandheads occur. The $\rm ^{12}C/^{13}C$ ratio is
constrained to $\rm ^{12}C/^{13}C= 20^{+12}_{-7}$ from ten $^{13}$CH A-X
lines in the region 4230-4250\,{\AA}. This $\rm ^{12}C/^{13}C$ ratio is
comparable with that observed in other metal-poor red giants that have
not yet experienced extra-mixing events arising at the red giant branch
(RGB) bump \citep{Spiteetal:2006}, confirming that CS~29497-004 sits on
the lower RGB (below the RGB bump).
 
The nitrogen abundance is derived from the violet NH band $\rm A^{3}
\Pi_{i} - X^{3} \Sigma^{-}$ at 3360\,{\AA}, and yields $\rm\log\,
\epsilon\,(N)= 5.35\,\pm\,0.1$ (see Fig. \ref{Fig:NHband}). The CN
molecule (3875-3900 \AA B-X 0-0 bandhead) is not detectable in
CS~29497-004, in agreement with the synthetic spectrum computed using the
C and N abundances derived from CH and NH, respectively. 

The OH lines in the UV portion of the spectrum permit measurements of the oxygen
abundance, based on twelve lines around between 3120\,{\AA} and 3180\,{\AA}.
The resulting oxygen abundance is $\rm\log\,\epsilon\,(O)=6.60\,\pm\,
0.1$. This oxygen abundance ($\rm\log\,\epsilon\,(O)=6.60$) was used to derive the C and N abundances above, 
and is used throughout the paper for all other synthesis, unless specified otherwise.
However, we find that there is a dependence of the abundance on
the OH line strength, the weaker lines giving lower oxygen abundances.
This illustrates that in EMP stars, molecules (not only OH, but also CH,
NH, and CN) that form high up in the stellar atmosphere are sensitive to 3D
effects, and probably also to NLTE effects, as emphasized by
\citet{Asplund:2005}, and more specifically for giants by \citet{Colletetal:2007}. In the latter work, a
dependence on the 1D to 3D correction on the excitation potential (E.P.)
of the OH lines is expected,  and this is also confirmed with a different 3D code by \citet{Dobrovolskasetal:2013}. 
This could (at least partly) explain the variations
found here between the weak and strong O lines. All lines measured here around
3120-3180{\AA} have E.P. in the range 0.1 to 0.8\,eV, while the weak OH
lines found around 3300\,{\AA} have E.P. around 1.5\,eV. For
\object{CS~29497-004}, a significantly lower O abundance of $\rm\log\,
\epsilon(O)=6.30\,\pm\, 0.15$ is found when evaluated from these
particular weak lines (see Fig. \ref{Fig:OHband}).

\begin{figure}[htbp]
  \centering
  \includegraphics[width=\hsize]{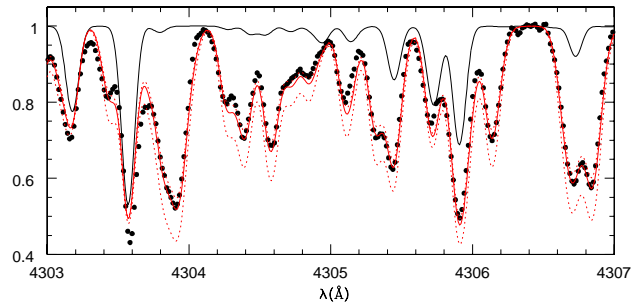}
  \caption{ Observed (dots) and synthetic spectra (lines) around the 
CH A-X bandhead ($G$-band) at 4300\,{\AA}, where the solid and dashed red lines are computed 
with $\log\epsilon\rm(C)=5.89, 5.94$, and $6.09$, and the solid black line is a synthetic 
spectrum without CH molecular lines.}\label{Fig:Gband}
\end{figure}

\begin{figure}[htbp]
  \centering
  \includegraphics[width=\hsize]{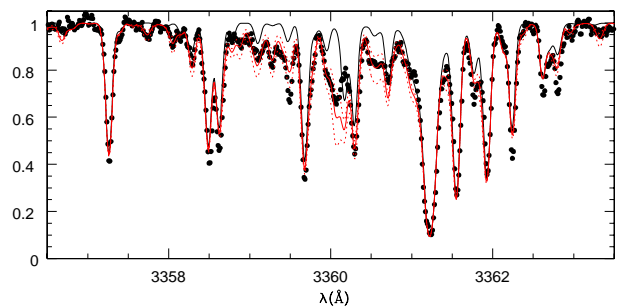}
  \caption{ Observed (dots) and synthetic spectra (lines) around the 
NH bandhead at 3360\,{\AA}, where the solid and dashed red lines are computed 
with $\log\epsilon\rm(N)=5.15, 5.35$, and $5.55$, and the solid black line is a synthetic 
spectrum without NH molecular lines.}\label{Fig:NHband}
\end{figure}

\begin{figure}[htbp]
  \centering
  \includegraphics[width=\hsize]{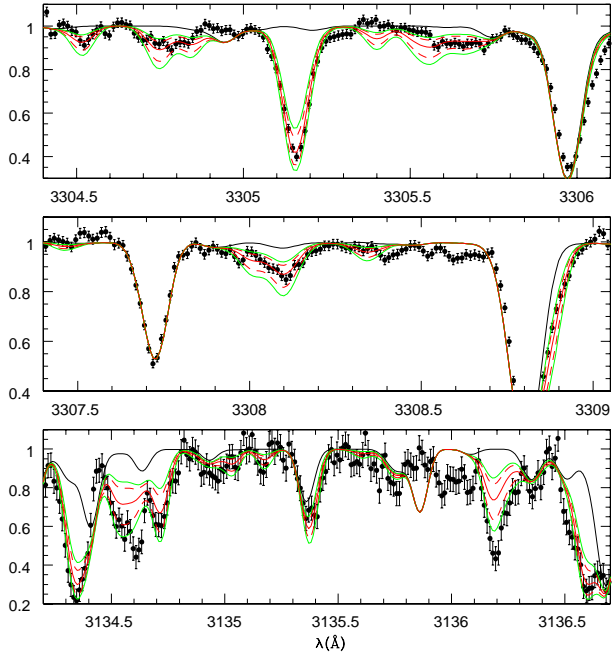}
  \caption{ Observed (dots) and synthetic spectra (lines) around  
various OH lines, where the solid and dashed red and green
lines are computed with $\log\epsilon\rm(O)=6.00,6.10, 6.30, 6.50$, and
$6.60$, and the solid black line is a synthetic 
spectrum without OH molecular lines.}\label{Fig:OHband}
\end{figure}

\subsection{Neutron-capture elements}\label{Sect:HeavySynthesis}

The abundances for individual lines of the neutron-capture elements are
given in Table~\ref{Tab:LinelistHeavy}, together with the method that
was used to derive them. When the lines were sufficiently isolated, and
no hyperfine structure is present, the abundances were deduced from
equivalent widths, while line-profile fitting with spectrum synthesis
was performed in all other cases, including all the weakest lines.

Below, we comment on the detection of special lines, element by element.

\subsubsection{Light $r$-process elements: $39 < \mathrm{Z} \le 48$}

{\em Molybdenum:}
This species has only one weak line detectable from the ground, at 3864.10\,{\AA}. 
In the carbon-enhanced $r$-II star \object{CS~22892-052},
\citet{Snedenetal:2003} remark that, in addition to
being very weak, the Mo line sits quite close
to a CN line, which could account for at least a part of the absorption at the Mo wavelength.
Fortunately, \object{CS~29497-004} has extremely low C and N content, and no
CN absorption is visible in the spectrum. The Mo line therefore
appears clean in our spectrum, and is very clearly detected, as illustrated in Fig.~\ref{Fig:Mo}. 

%For comparison, in CS~22892-052, the \ion{Fe}{I}+CN
%absorption at 3864.3\,{\AA} is deeper than the Mo+CN line, owing to the strong
%CN content of this star, whereas in CS~29497-004, the Mo feature
%should be essentially free of CN.

\begin{figure}[htbp]
  \centering
  \includegraphics[width=\hsize]{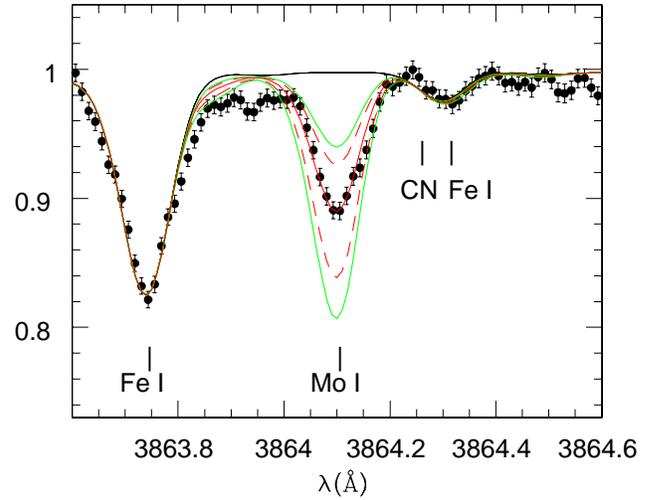}
  \caption{ Observed (dots with error bars) and synthetic spectra (lines) around the 
Mo line at 3864.10\,{\AA} for abundances of $\rm
\log\,\epsilon(Mo)\,=\,-0.15,-0.05,0.05,0.15,0.25,0.35,0.45$, and no
Mo.}\label{Fig:Mo}
\end{figure}

We searched for cadmium (Cd at 3261.0\,{\AA}) and tin (Sn at 3262.3
and 3801.0\,{\AA}), but these transitions were too weak and too blended
to enable detection. 

\subsubsection{Second $r$-process peak elements: $56 <\mathrm{Z} \le 72$}\label{sec_2ndpeak_detection}

{\em Lutetium:} 
The lutetium abundance in CS~29497-004 was measured with the \ion{Lu}{II}
3397\,{\AA} line (see Fig. \ref{Fig:Lu}), and further checked using
the very weak \ion{Lu}{II} features at 3472.45 and 3554.39\,{\AA}.
Despite the fact that the 3397\,{\AA} line is in
the wing of an \ion{Fe}{I} line, the abundance of Lu is 
well-constrained to $\log\,\epsilon(\mathrm{Lu})\,=\,-0.72\,\pm0.1$, 
and the two other weak features agree with this measurement within
their (larger) measurement uncertainties.

\begin{figure}[htbp]
  \centering
  \includegraphics[width=\hsize]{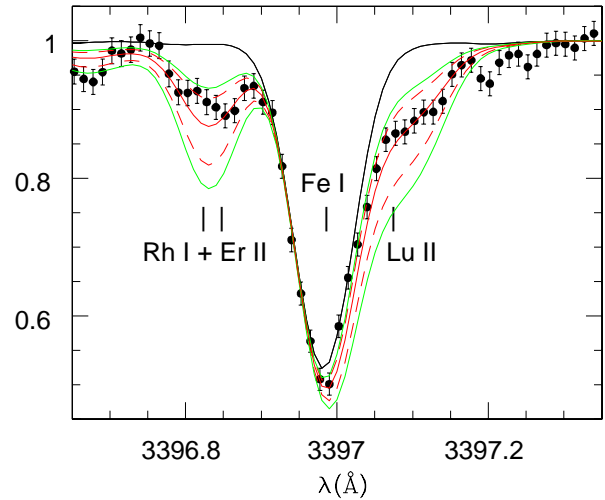}
  \caption{ Observed (dots with error bars) and synthetic spectra (lines) around the 
Lu line at 3397\,{\AA} for Lu abundances of $\rm
\log\,\epsilon(Lu)\,=\,-1.02,-0.92,-0.72,-0.52$, and $-0.42$ as red and green
lines, and no Lu (black line). Despite the fact that this line is in
the wing of an \ion{Fe}{I} line, the abundance of Lu is well-constrained.}\label{Fig:Lu} 
\end{figure}

\subsubsection{Third $r$-process peak elements and the actinides: $72 < \mathrm{Z}
  \le 92$}\label{detection 3rd peak}

{\em Platinum:} From the ground, the only metal-poor star in which
platinum has been detected so far is the $r$-II star
\object{CS~22892-052} \citep{Snedenetal:2003}, from two lines at 3301.9
and 3315.0\,{\AA}. Pt detections were obtained for a handful of other
metal-poor stars from stronger lines in the near-UV around 2650 and 2900\,
{\AA} with HST/STIS \citep[e.g.,][]{Westinetal:2000, Cowanetal:2002,
Barbuyetal:2011}.

The 3315.0\,{\AA} \ion{Pt}{I} line is not detected in our spectrum (the
continuum in this region is not sufficiently clean to permit detection
of this line, which has an equivalenth width of only a few m{\AA}). The
3301.87\,{\AA} feature is stronger and clearly visible, but an OH line
is predicted to fall right on top of it (at 3301.85\,{\AA}), accounting
for $\sim$40\% of the total absorption at the Pt wavelength. An oxygen
abundance of $\rm\log\,\epsilon(O)=6.30\,\pm\,0.15$ has been used to
model OH lines in the wavelength region around the Pt line at 3301{\AA}
(see discussion in section \ref{Sect:MolSynthesis}). The uncertainty in
the OH feature translates into a 0.2\,dex uncertainty in the Pt
abundance for \object{CS~29497-004} (see Fig. \ref{Fig:Pt}). We further
checked that in the $r$-process-poor star \object{HD~122563} ([Fe/H] =
$-2.71$, $T_{\mathrm{eff}}$ = 4625~K, $\log g$ = 1.4; \citealt{Cayreletal:2004}),
where no Pt is visible, the relative strengths of the OH lines
at 3301.85\,{\AA} (blending the Pt line), 3305.7\,{\AA}, and 3308.3\,{\AA}, (employed
to set the O abundance) were succesfuly reproduced by a synthetic
spectrum using the same OH linelist. This strengthens our measurement of
the Pt abundance for \object{CS~29497-004}. We further note that, for
\object{CS~22892-052}, where Pt is detected both at 3301.87\,{\AA} and through the
cleaner and stronger UV line at 2929.82\,{\AA} \citep{Snedenetal:2003},
both abundances agree well (within $0.15$\, dex).

%We however note that in CS22892-052, the deduced Pt
%abundance from this line is much larger. However, the same OH blend is
%to be expected in this star, and the resulting Pt abundance is
%therefore also dependent on the assumed O abundance and OH line
%strength. 

\begin{figure}[htbp]
  \centering
  \includegraphics[width=\hsize]{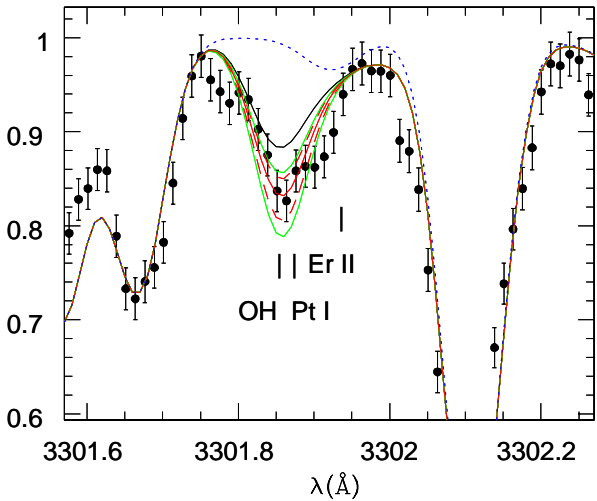}
  \caption{ Observed (dots with error bars) and synthetic spectra
 (lines) around the  Pt line at 3301.87\,{\AA} for abundances of
 $\rm\log\,\epsilon(Pt)\,=\,-0.3,-0.2,0.0,+0.2$, and $+0.3$ (green and red
 full and dashed curves). In addition, a synthesis with no Pt (black
 full line) shows the prominent OH blend, while the upper dotted blue
 curve shows a synthesis without Pt and OH, revealing a faint
 \ion{Er}{II} line in the red wing of the feature, which has no impact
 on the Pt abundance determination.}\label{Fig:Pt}
\end{figure}

{\em Lead:}
Lead is on the edge of detection for \object{CS~29497-004} with our current
spectra. The 3665\,{\AA} line is not detected, while the 4057.8\,{\AA}
feature is exceedingly weak, and in the vicinity of a $^{12}$CH feature
(from the B-X system), only allowing an upper limit on the Pb abundance
to be derived. A conservative upper limit of $\rm\log\,\epsilon(Pb)\,<\,
+0.25$ is derived from $\chi ^2$ minimisation (at a 90\% confidence
level), independent of the assumed carbon abundance for the nearby
$^{12}$CH transition (see Fig. \ref{Fig:Pb}). The actual minimum of the
$\chi ^2$ is reached for $\rm\log\,\epsilon(Pb)\,=\,-0.25$. Both values
are reported in Table~\ref{Tab:AbundancesHeavy}.

\begin{figure*}[htbp]
  \centering
\includegraphics[angle=-90,width=\hsize]{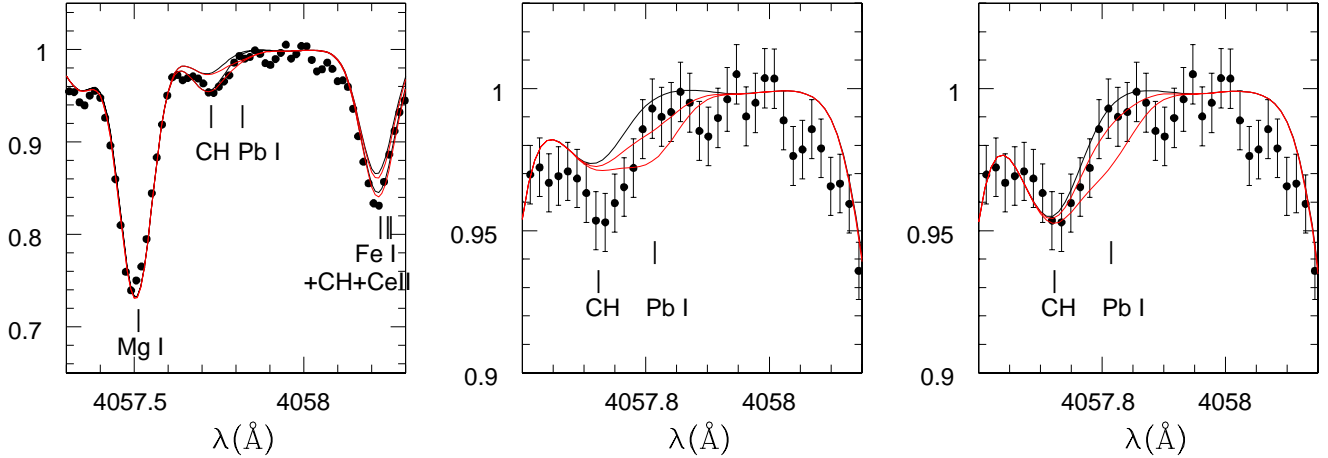}
\caption{ Observed (dots with error bars) and synthetic spectra
(lines) around the  Pb line at 4057.8\,{\AA}. This Pb line lies in
the red wing of a weak $^{12}$CH feature (from the B-X system), which
proved to be underestimated with the carbon abundance derived from the
G-band ($^{12}$CH A-X system), as shown in the middle panel.
Two hypotheses for the abundances of carbon have therefore been
considered here: (1) The nominal abundance $\rm\log\,\epsilon(C)\,=\,5.94$ (middle
panel), and (2) An abundance $\rm\log\,\epsilon(C)\,=\,6.20$, which
best fits the CH 4057.7\,{\AA} line (right panel). The left panel shows together the 
best-fit values for Pb (red curves) and the synthesis with no Pb (black curves)
in these two cases. {\em (middle panel)} Syntheses are shown for
$\rm\log\,\epsilon(Pb)\,=\,-0.15\, and +0.25$ (respectively, the best-fit
and a conservative upper limit, shown in red), and no Pb, assuming the nominal
C abundance. {\em (right panel)} $\rm\log\,\epsilon(Pb)\,=\,-0.25\, and
+0.25$ (respectively,
the best-fit and a conservative upper limit, shown in red), and no Pb,
assuming $\rm\log\,\epsilon(C)\,=\,6.20$.}\label{Fig:Pb} 
\end{figure*}

{\em Thorium:} Owing to the high levels of $r$-process elements in
\object{CS~29497-004}, the radioactive element Th is measured from a total of
six lines, all of which have accurate oscillator strength data from
\citet{Nilssonetal:2002b}. The abundances from these lines all agree
with one another within the errors; four of these six detections are illustrated in
Fig.~\ref{Fig:Th}. Note in particular that, thanks to the low C content
of this star, the commonly employed 4019\,{\AA} line is virtually free of the
$^{13}$CH contamination from which it suffers in C-enhanced stars.

\begin{figure*}[htbp]
  \centering
  \includegraphics[width=\hsize]{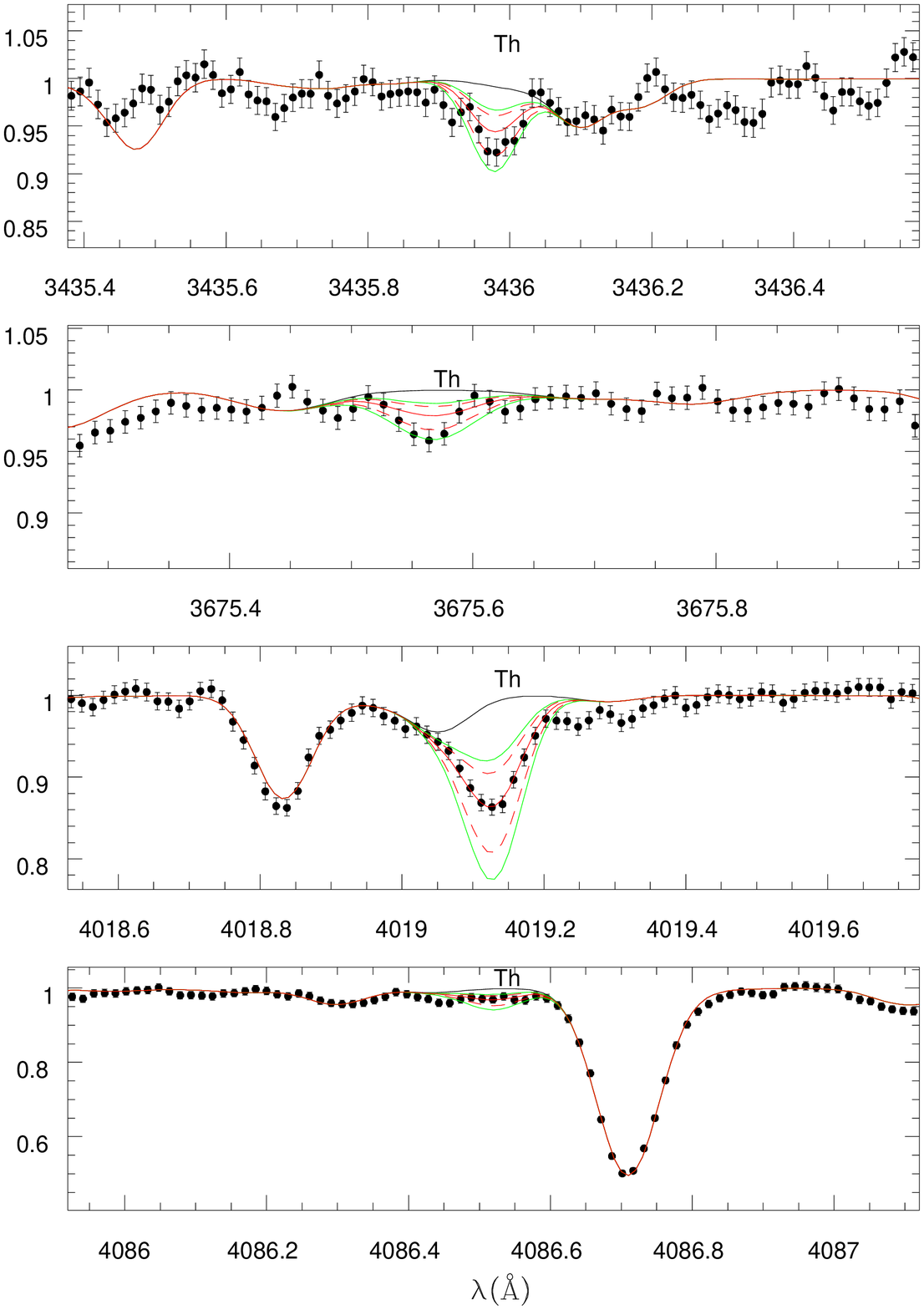}
  \caption{ Observed (dots with error bars) and synthetic spectra
    (lines) around four Th lines for abundances of
    $\rm\log\,\epsilon(Th)\,=\,-1.55,-1.45,-1.25,-1.05,-0.95$,
    and no Th.}\label{Fig:Th} 
\end{figure*}

{\em Uranium:} The radioactive element U is very difficult to detect,
because most of its transitions are weak and blended. The U transition
detected for \object{CS~29497-004} (at 3859.57\,{\AA}) lies in the wing of a \ion{Fe}{I} line at
3860\,{\AA}.  To date, uranium has only been unambiguously detected in
two EMP stars: \object{CS~31082-001} \citep{Cayreletal:2001,
Hilletal:2002} and \object{HE~1523-0901} \citep{Frebeletal:2007}. 
With a temperature of 5013\,K, \object{CS~29497-004}
is slightly warmer than both of these stars (with $T_{\mathrm{eff}}$
of 4825 and 4630\,K respectively) and lower on the RGB ($\log g = 2.05$
instead of 1.55 and 1.00, respectively, for the other two stars). This
results in the U line being intrinsically weaker (and the wing of the
Fe~I line being relatively stronger) for \object{CS~29497-004}.
Nevertheless, we could detect uranium for
\object{CS~29497-004} at a level of $\rm\log\,\epsilon(U)
\,=\,-2.20\,\pm\,0.3$ (see Fig.~\ref{Fig:U}). We have carefully examined
potential sources of uncertainties that most affect the measurement of
uranium: photon noise, and continuum placement and  \ion{Fe}{I}-wing
modelling.

\begin{itemize}

\item Photon noise: The expected noise in the spectrum was propagated throughout the
  reduction chain from the raw image to the final co-added
  spectrum; error bars in Fig.~\ref{Fig:U} reflect this noise. Since the
  U-line region is covered by 28 spectra, we also estimated the
  standard deviation around the weighted average of all spectra for
  each pixel. This standard deviation is extremely close to the
  error propagated along the reduction chain, demonstrating that our
  estimate of the noise is indeed valid. According to these various
  methods of estimating the noise, total $S/N$ ratios of 170 to
  200 are measured per 0.012\,{\AA} pixel in this region. Of these individual spectra,
  22 were taken with the B436 setting and 6 from the B437 setting. We
  also compared, in the U-line  region, the spectra co-added separately 
  for the two settings. The line profiles are indistinguishable within
  their respective error bars. 

\item Continuum placement and \ion{Fe}{I} wing: The placement of the continuum is of course an
  important factor in the fit of a line as weak as the U
  transition. It was determined in regions clean of lines surrounding
  the region of interest -- around  3859.0\,{\AA} and
  $3860.4-3860.6\,${\AA}. However, the continuum placement, by itself, is
  not the primary limiting factor that influences measurement of the uranium
  line. The blue wing of the $3859.9\,$\AA\ \ion{Fe}{I} line, at the
  temperature of \object{CS~29497-004}, is the dominant absorption
  feature immediately to the red of the U line. With the adopted continuum, we adjusted the
  line broadening of the \ion{Fe}{I} line to ensure that its red wing
  (which is essentially free of blends) would fit the observed spectrum
  as closely as possible. The combined uncertainties of this and the
  continuum placement amount to a $\sim 0.2\,$dex uncertainty on the
  measured uranium abundance.
\end{itemize}

\begin{figure*}[htbp]
  \centering
  \includegraphics[width=\hsize]{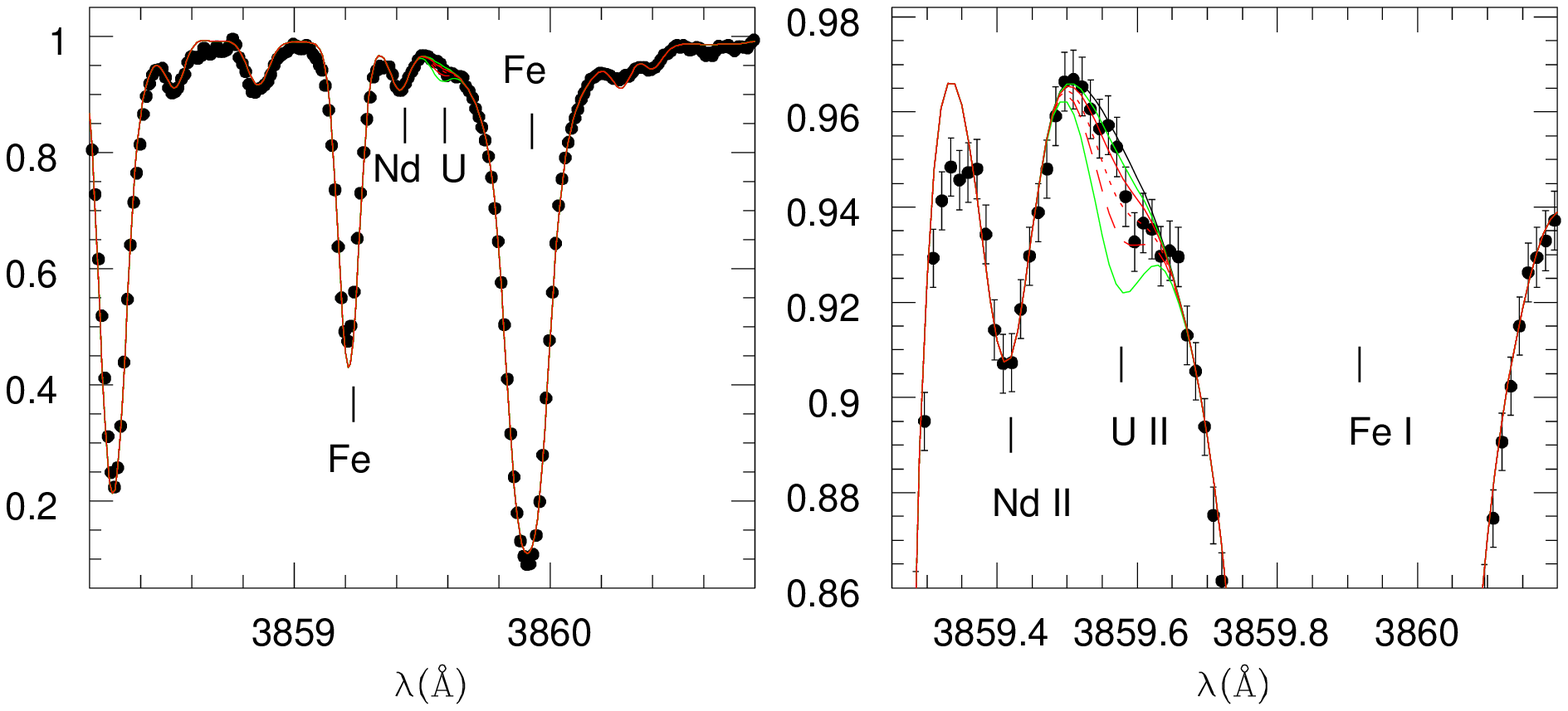}
  \caption{ Observed (dots with error bars) and synthetic spectra (lines) around the 
uranium line at 3859.57\,{\AA} for abundances of $\rm\log\,\epsilon(U)\,
=\,-2.50 -2.20 -1.90 -1.50$, and no U. }\label{Fig:U}
\end{figure*}

\section{Results}\label{Sect:Results}

\subsection{C through Zn}

The abundances of all elements from Mg through Zn were measured using
equivalent widths, and are reported in Table~\ref{Tab:AbundancesLight}. 
\object{CS 29497-004} does not stand out in any particular way
compared to other giants of similar low metallicity in the Galactic halo, as
probed for example by \cite{Cayreletal:2004}, \cite{HERESpaperII}, \cite{Laietal:2008}, or
\citet{RoedereretalEMPS:2014}. It exhibits (in the LTE approximation) the same
overabundances of the $\alpha$-elements (Mg, Si, Ca, Ti) as well as Co, Zn, and
typical underabundances of Al, Cr, and Mn. We caution that these are LTE
abundances, and recall that NLTE and/or 3D effects can alter these results
\citep[see, e.g., ][for a review]{Asplund:2005}. 
In the case of Al, for example, large NLTE departures are expected:
\cite{Andrievskyetal:2008} have shown that NLTE corrections for the
lines used in EMP stars are strong, and rely sensitively on the stellar
atmospheric parameters. For giants, the hotter and the more metal-poor
the star, the stronger the NLTE effects, while the more luminous the
giant, the weaker the NLTE corrections. For a giant of
$T_{\mathrm{eff}}$ = 5000~K, $\log g$ = 2, [Fe/H] = $-3$ (similar to
\object{CS~29497-004}), the expected NLTE correction for Al is on the
order of +0.7~dex, which would bring the aluminum abundance of
\object{CS~29497-004} back to a Solar [Al/Fe] ratio.

The abundances of C, N, and O for \object{CS~29497-004} are also
reported in Table \ref{Tab:AbundancesLight}. We emphasise that these
abundances are computed in 1D LTE, and could be subject to relatively
large revisions once the 3D and NLTE effects are taken into account. As
an illustration, the 1D to 3D corrections for a $T_{\mathrm{eff}}$ =
5050~K giant of similar gravity to \object{CS~29497-004} amount to about
$-1.1$\,dex for OH, $-0.7$\,dex for CH and $-0.9$\,dex for NH, according
to \cite{Colletetal:2007}. Under these assumptions, the 3D CNO abundances
of \object{CS~29497-004} would be all sub-Solar, $\mathrm{[C/Fe]} = -0.3$,
$\mathrm{[N/Fe]} = -0.5$, and $\mathrm{[O/Fe]}=-0.3$, respectively.
Again, the 1D LTE C, N and O abundances found here are in agreement with
those of giants in the sample of \cite{Spiteetal:2005}. More
specifically, both the C and N abundances, as well as the isotopic
ratio $\rm^{12}C/^{13}C$ of \object{CS~29497-004}, are as expected for
its luminosity (below the RGB bump), similar to the {\it unmixed} giants
of \cite{Spiteetal:2005} and \cite{Spiteetal:2006}. These stars have all
passed the first dredge-up, but have not (yet) experienced strong extra
mixing of internal CNO-cycled material that is triggered by the fading
of the molecular-weight barrier at the RGB bump. 

In summary, the lighter element abundances from C to Zn in
\object{CS~29497-004} are indistinguishable from other EMP giant stars
of the same metallicity. 

\begin{table}[htbp]
  \caption{\label{Tab:AbundancesLight} Abundances of the elements
    up to the iron peak for \object{CS~29497-004}. The Solar C, N, and O
    abundances were taken from the 1D MARCS analysis of
    \citet{Asplundetal:2005a}, while all other Solar abundances are
    from \cite{Grevesse/Sauval:2000}.}
  \centering
  \begin{tabular}{rlrrrcr}\hline\hline
    $Z$ & Species & \multicolumn{1}{l}{$\log\epsilon_{\odot}$} & 
  \multicolumn{1}{l}{$\log\epsilon$} & $\sigma$ & \multicolumn{1}{l}{$N_{\mathrm{lines}}$}
 &\multicolumn{1}{l}{[X/Fe]} \\\hline
 $6$&  $^{12}\mathrm{C}/^{13}\mathrm{C}$ & 80   &  20  & $^{+12}_{-7}$\\
 $6$&  C~(CH)       & $8.42$ &  $5.94$ & $0.05$ &      & $ 0.37$  \\
 $7$&  N~(NH)       & $7.82$ &  $5.35$ & $0.10$ &      & $ 0.38$  \\
% $7$&  N~(CN)       & $7.78$ &  $5.54$ &       &      & $0.02$ \\
 $8$&  O~(OH)       & $8.72$ &  $6.60$ & $0.10$ &      & $ 0.73$ \\ %strong lines at 3100A
% $8$&  O~(OH)       & $8.66$ &  $6.30$ & $0.10$&      & $ 0.49$ \\ %weak lines around 3300A
 $12$& \ion{Mg}{I}  & $7.58$ &  $5.11$ & $0.27$ &  $5$ & $ 0.38$  \\
 $13$& \ion{Al}{I}  & $6.47$ &  $2.81$ & $0.00$ &  $1$ & $-0.81$  \\
 $14$& \ion{Si}{I}  & $7.55$ &  $5.16$ & $0.11$ &  $2$ & $ 0.46$  \\
 $20$& \ion{Ca}{I}  & $6.36$ &  $3.81$ & $0.13$ &  $5$ & $ 0.30$  \\
 $21$& \ion{Sc}{II} & $3.17$ &  $0.42$ & $0.07$ &  $4$ & $ 0.10$  \\
 $22$& \ion{Ti}{I}  & $5.02$ &  $2.40$ & $0.03$ &  $4$ & $ 0.23$  \\
 $22$& \ion{Ti}{II} & $5.02$ &  $2.45$ & $0.09$ & $22$ & $ 0.28$  \\
 $24$& \ion{Cr}{I}  & $5.67$ &  $2.50$ & $0.04$ &  $3$ & $-0.32$  \\
 $25$& \ion{Mn}{I}  & $5.39$ &  $2.31$ & $0.12$ &  $5$ & $-0.23$  \\
 $26$& \ion{Fe}{I}  & $7.50$ &  $4.65$ & $0.12$ & $77$ & $ 0.00$  \\
 $26$& \ion{Fe}{II} & $7.50$ &  $4.66$ & $0.12$ & $13$ & $ 0.01$  \\
 $27$& \ion{Co}{I}  & $4.92$ &  $2.45$ & $0.05$ &  $4$ & $ 0.38$  \\
 $28$& \ion{Ni}{I}  & $6.25$ &  $3.54$ & $0.01$ &  $2$ & $ 0.14$  \\
 $30$& \ion{Zn}{I}  & $4.60$ &  $2.04$ & $0.00$ &  $2$ & $ 0.29$  \\
\hline
  \end{tabular}
\end{table}

\subsection{Neutron-capture elements}

The final neutron-capture element abundances for \object{CS~29497--004}
are listed in Table~\ref{Tab:AbundancesHeavy}, and plotted as a function
of atomic number in Fig.~\ref{Fig:comp3stars}. The table contains, in
addition to the mean abundance for each element, the dispersion around
the mean ($\sigma$), the number of lines used in the analysis
($N_{\mathrm{lines}}$), and the total error on the [Th/X] ratio
($\sigma_t$). This last error estimate takes into account both the total
measurement error and the effect of the stellar parameter uncertainties
on the [Th/X] ratio ($\sigma_m$), combining both sources in quadrature.
The total measurement error itself is computed as the quadratic sum of
the errors on element X and Th, estimated as
$\rm max[\sigma(X)/\sqrt{N(X)},\sigma_{\rm expected}(X)]$, where 
$\sigma_{\rm expected}$ is the propagated individual-line measurement error,  and  
$\sigma(Th)/\sqrt{N(\rm Th)}$ respectively.  

\begin{table}[htbp]
  \caption{\label{Tab:AbundancesHeavy} Abundances of the 
    neutron-capture elements for \object{CS~29497-004}. 
Column $\sigma_m$ reports errors on the [Th/X] ratios propagated from stellar parameter uncertainties, and
 $\sigma_t$ is the total error on the [Th/X] ratio (combining the
    observational uncertainty and $\sigma_{m}$).}
  \centering
  \begin{tabular}{rlrrrcrr}\hline\hline
    $Z$ & Species & \multicolumn{1}{l}{$\log\epsilon_{\odot}$} & 
    \multicolumn{1}{l}{$\log\epsilon$} & $\sigma$ & \multicolumn{1}{l}{$N_{\mathrm{lines}}$} &
    \multicolumn{1}{l}{$\sigma_m$} & \multicolumn{1}{l}{$\sigma_{t}$} 
%    \multicolumn{1}{l}{$\sigma_{\mathrm model}$([X/Th])} & \multicolumn{1}{l}{$\sigma_{\mathrm tot}$([X/Th])} 
\\\hline
$38$ & \ion{Sr}{II} & $2.97$ & $ 0.76$ & $0.11$ & $ 4$ & $0.07$ & $0.11$  \\  
$39$ &  \ion{Y}{II} & $2.24$ & $-0.04$ & $0.11$ & $13$ & $0.06$ & $0.08$  \\ 
$40$ & \ion{Zr}{II} & $2.60$ & $ 0.63$ & $0.14$ & $27$ & $0.04$ & $0.07$  \\ 
$41$ & \ion{Nb}{II} & $1.42$ & $-0.33$ & $0.00$ & $ 1$ & $0.01$ & $0.07$  \\ 
$42$ & \ion{Mo}{I}  & $1.92$ & $ 0.15$ & $0.00$ & $ 1$ & $0.13$ & $0.14$  \\ 
$44$ & \ion{Ru}{I}  & $1.84$ & $ 0.61$ & $0.04$ & $ 5$ & $0.13$ & $0.15$  \\ 
$45$ & \ion{Rh}{I}  & $1.12$ & $-0.18$ & $0.12$ & $ 2$ & $0.13$ & $0.16$  \\ 
$46$ & \ion{Pd}{I}  & $1.69$ & $ 0.25$ & $0.23$ & $ 5$ & $0.13$ & $0.18$  \\ 
$47$ & \ion{Ag}{I}  & $1.24$ & $-0.42$ & $0.11$ & $ 2$ & $0.13$ & $0.17$  \\ 
$56$ & \ion{Ba}{II} & $2.13$ & $ 0.35$ & $0.17$ & $ 5$ & $0.08$ & $0.12$  \\ 
$57$ & \ion{La}{II} & $1.17$ & $-0.38$ & $0.14$ & $13$ & $0.05$ & $0.08$  \\ 
$58$ & \ion{Ce}{II} & $1.58$ & $-0.19$ & $0.12$ & $31$ & $0.02$ & $0.06$  \\ 
$59$ & \ion{Pr}{II} & $0.71$ & $-0.67$ & $0.04$ & $ 6$ & $0.02$ & $0.06$  \\ 
$60$ & \ion{Nd}{II} & $1.50$ & $-0.02$ & $0.09$ & $33$ & $0.02$ & $0.07$  \\ 
%$62$ & \ion{Sm}{II} & $1.01$ & $-0.26$ & $0.17$ & $17$ & $0.01$ & $0.07$  \\ % old loggf
$62$ & \ion{Sm}{II} & $1.01$ & $-0.30$ & $0.11$ & $17$ & $0.01$ & $0.06$  \\ % corrected loggf by Lawler etal. 2006
$63$ & \ion{Eu}{II} & $0.51$ & $-0.66$ & $0.04$ & $ 7$ & $0.02$ & $0.06$  \\ 
$64$ & \ion{Gd}{II} & $1.12$ & $-0.31$ & $0.12$ & $20$ & $0.03$ & $0.07$  \\ 
$65$ & \ion{Tb}{II} & $0.35$ & $-1.08$ & $0.07$ & $ 9$ & $0.02$ & $0.06$  \\ 
$66$ & \ion{Dy}{II} & $1.14$ & $-0.08$ & $0.07$ & $21$ & $0.03$ & $0.07$  \\ 
$67$ & \ion{Ho}{II} & $0.51$ & $-0.76$ & $0.04$ & $ 3$ & $0.15$ & $0.17$  \\ 
%$68$ & \ion{Er}{II} & $0.93$ & $-0.23$ & $0.15$ & $ 9$ & $0.08$ & $0.11$  \\  %old loggf
$68$ & \ion{Er}{II} & $0.93$ & $-0.29$ & $0.11$ & $ 9$ & $0.08$ & $0.11$  \\ % corrected loggf by Lawler etal. 2008
$69$ & \ion{Tm}{II} & $0.15$ & $-1.18$ & $0.05$ & $ 6$ & $0.05$ & $0.08$  \\ 
$70$ & \ion{Yb}{II} & $1.08$ & $-0.22$ & $0.06$ & $ 2$ & $0.17$ & $0.20$  \\ 
$71$ & \ion{Lu}{II} & $0.06$ & $-0.72$ & $0.00$ & $ 1$ & $0.02$ & $0.11$  \\ 
$72$ & \ion{Hf}{II} & $0.88$ & $-0.64$ & $0.07$ & $ 9$ & $0.02$ & $0.07$  \\ %new list and loggfs 
$76$ & \ion{Os}{I}  & $1.45$ & $ 0.41$ & $0.10$ & $ 5$ & $0.10$ & $0.12$  \\ 
$77$ & \ion{Ir}{I}  & $1.35$ & $ 0.23$ & $0.02$ & $ 2$ & $0.06$ & $0.09$  \\ 
$78$ & \ion{Pt}{I}  & $1.80$ & $ 0.00$ & $0.00$ & $ 1$ & $0.06$ & $0.21$  \\ 
$82$ & \ion{Pb}{I}  & $1.95$ & $-0.25$ & $0.00$ & $ 1$ & $0.10$ & $0.32$  \\ %best fit 
$82$ & \ion{Pb}{I}  & $1.95$ & $<0.25$ & $0.00$ & $ 1$ & $0.10$ & $0.32$  \\ %conservative limit  
$90$ & \ion{Th}{II} & $0.09$ & $-1.16$ & $0.13$ & $ 6$ & $0.00$ & $0.08$  \\ 
$92$ & \ion{ U}{II} & $0.50$ & $-2.20$ & $0.00$ & $ 1$ & $0.01$ & $0.32$  \\ 
\hline
  \end{tabular}
\end{table}

\begin{figure*}[htbp]
  \centering
  \includegraphics[width=\hsize]{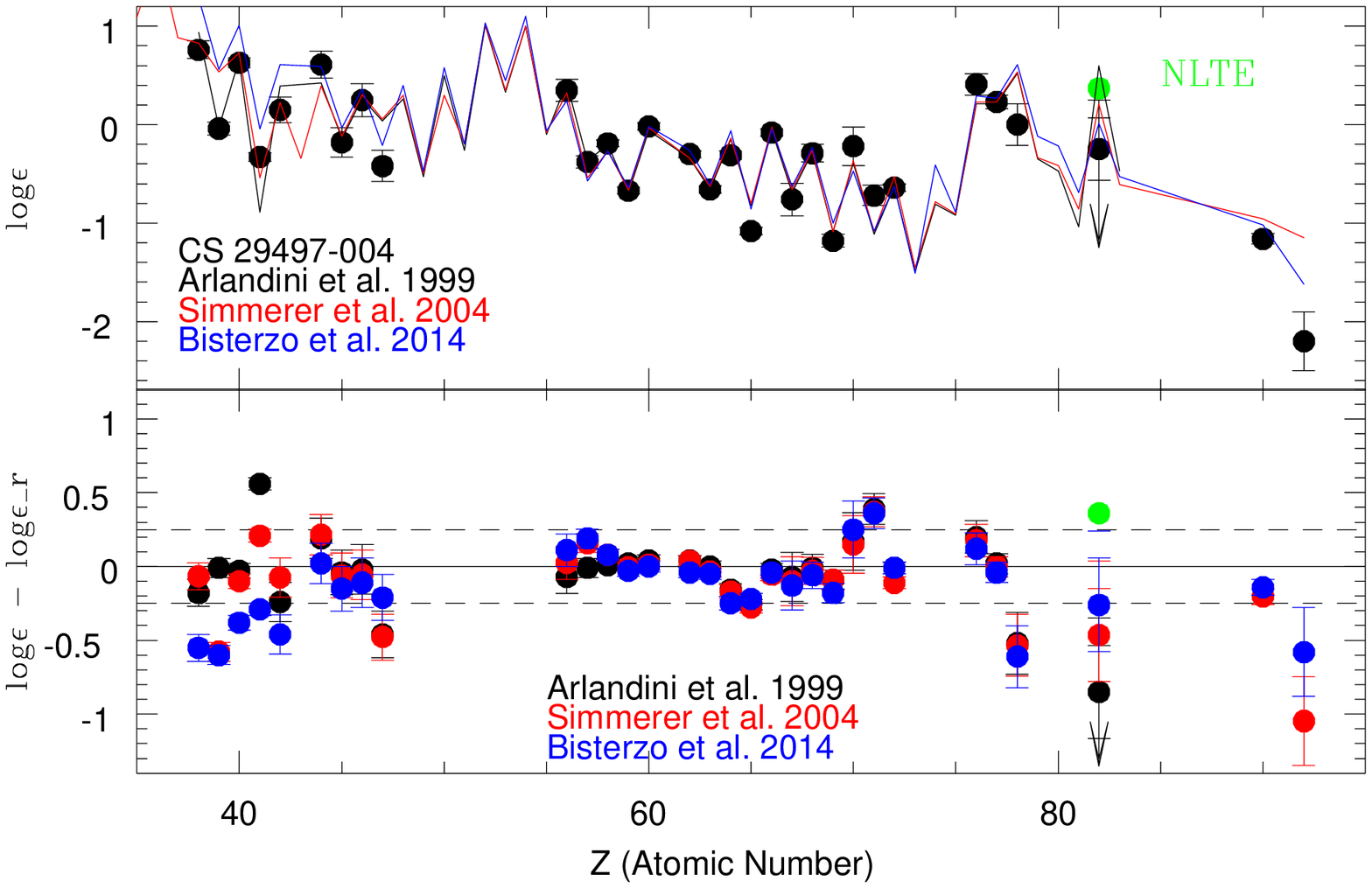}
  \caption{{\em Upper panel}: Neutron-capture element abundances for
  \object{CS~29497-004} plotted against atomic number, and
  compared to the Solar $r$-process abundances, following the
  decompositions of \cite{Arlandinietal:1999} (black line), 
  \cite{Simmereretal:2004} (red line), and \cite{Bisterzoetal:2014}, using the \cite{Loddersetal:2009} 
  Solar composition (blue line).  {\em Lower panel}: Residuals of
  the abundances for \object{CS~29497-004} to the Solar $r$-process
  decompositions, normalised to the mean $r$-process element abundances
  with $55<\rm Z<73$. The green point in both panels corresponds to the Pb abundance 
  in \object{CS~29497-004} corrected for NLTE as of \citet{Mashonkinaetal:2012}.}\label{Fig:compth} 
\end{figure*}

\begin{figure*}[htbp]
  \centering
  \includegraphics[width=\hsize]{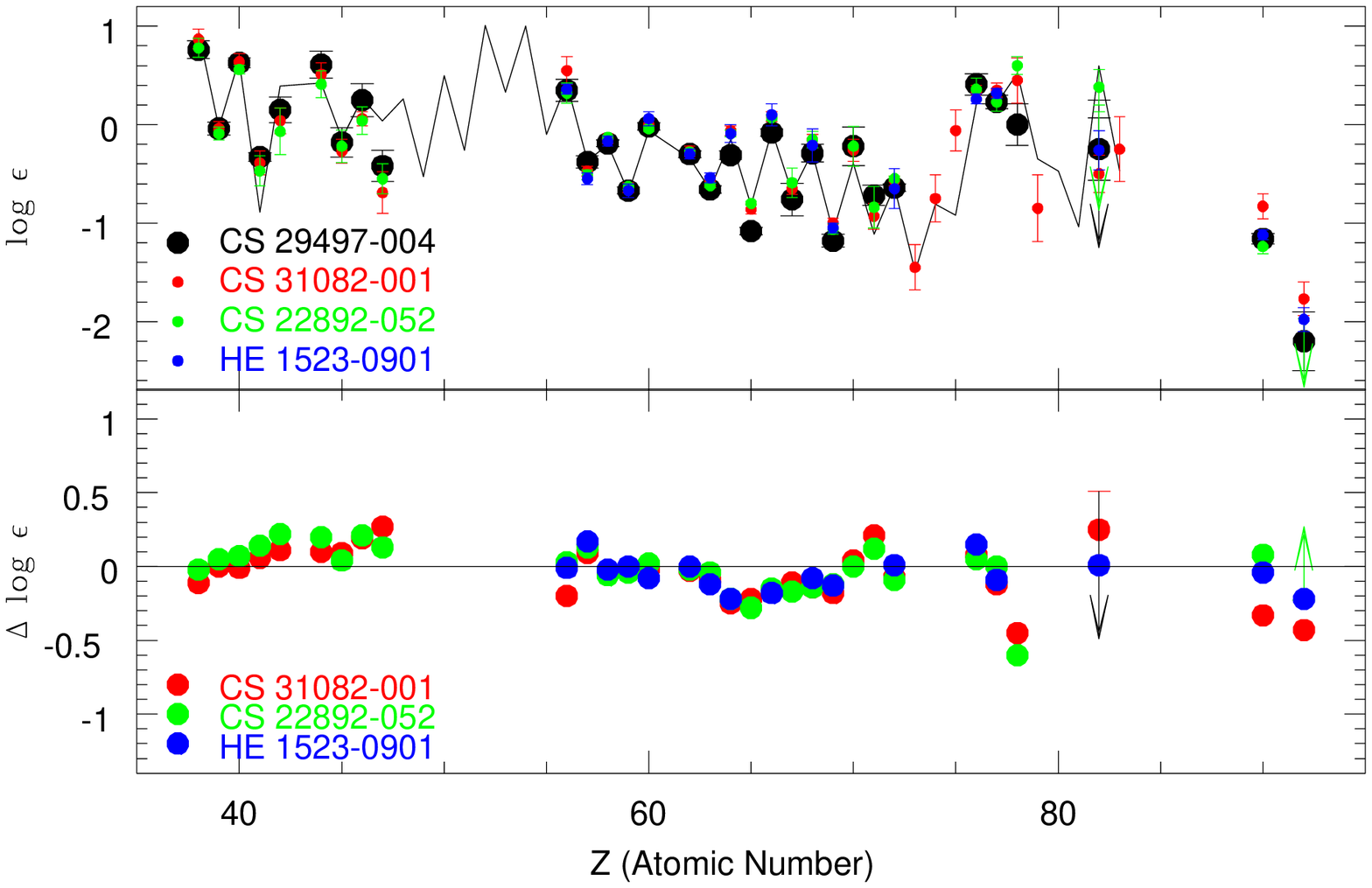}
  \caption{{\em Upper panel}: Neutron-capture element abundances for
  \object{CS~29497-004} (black), compared to those in other $r$-II stars (normalised at Ce): 
\object{CS~31082-001} \citep[red, ][]{Hilletal:2002, Snedenetal:2009, Barbuyetal:2011, SiqueiraMelloetal:2013}, 
\object{CS~22892-052} \citep[green, ][]{Snedenetal:2003, Snedenetal:2009}, and
\object{HE~1523-0901} \citep[blue, ][]{Frebeletal:2007}. All abundances are computed under the assumption of LTE. 
The Solar $r$-process decomposition of \cite{Arlandinietal:1999} is
  shown as a reference (black line).  
{\em Lower panel}: Abundance differences between \object{CS~29497-004}
  and the other $r$-II stars (normalised at Ce). }\label{Fig:comp3stars} 
\end{figure*}

\subsubsection{\object{CS~29497-004} versus Solar $r$-process abundances}

As illustrated in Fig.~\ref{Fig:compth}, the abundances of the
neutron-capture elements for \object{CS~29497-004} agree well with the Solar System
$r$-process distribution, as derived by \cite{Arlandinietal:1999}, 
\cite{Simmereretal:2004}, or \cite{Bisterzoetal:2014}, the latter using
an updated Solar composition by \cite{Loddersetal:2009} relying mostly
on meteoritic abundances for neutron-capture elements. The respective 
offsets of the \object{CS~29497-004} abundances with respect to these  Solar System $r$-process decomposition are:
$\rm \log \epsilon(X) - \log \epsilon(X_{r\odot}) = -1.15 \pm 0.14$ dex, 
$\rm \log \epsilon(X) - \log \epsilon(X_{r\odot}) = -1.12 \pm 0.15$ dex, and
$\rm \log \epsilon(X) - \log \epsilon(X_{r\odot}) = -1.10 \pm 0.16$ dex 
(these offsets were computed for all measured elements with $\rm 56\leq
Z\leq 72$). Taken together with the iron abundance of $\rm [Fe/H] =
-2.85$ for \object{CS~29497-004}, this confirms the very strong
$r$-process-element enhancement of this star, by some 1.7~dex, or a
factor of 50. 

The overall agreement of the observed heavy-element
abundances with the Solar $r$-process pattern is best 
for the second $r$-process peak elements ($55<\rm Z<75$).  Lutetium, at Z = $71$,
is the only element showing a discrepancy $>0.25$~dex, which may arise
partly from the observational error associated with the detection of this
element (see Sect.\ref{sec_2ndpeak_detection}).

The lighter $r$-process elements ($37<\rm Z<48$) are not as
well-reproduced by the Solar System decompositions. Although the scaled
pattern of \cite{Simmereretal:2004} is in slightly better agreement with
the observed abundances, the difference among the three Solar System
$r$-process decompositions indicates that the $r$-process fraction of these elements  suffer from more
uncertainties in this atomic mass range, perhaps not surprisingly given that their Solar 
System total abundance is dominated by the $s$-process contribution. 
The \cite{Bisterzoetal:2014} work uses the most up-to-date $s$-process prescriptions, 
so that the decomposition based on this work should be the most reliable of all presented here. 
In this framework, the first-peak $r$-process elements in \object{CS~29497-004} appear 
relatively less abundant than the second-peak $r$-process elements, as compared to the Solar System $r$-process
decomposition. Since all rII stars share very similar 
first-peak to second-peak $r$-process elements ratio (ex. Sr/Ba) ratios (see Sec.~\label{sec:comp_rII}), 
the same conclusion would hold for other rII stars too: the Solar System $r$-process has a higher first-peak to second-peak $r$-process elements 
than rII stars. 
Although it cannot be excluded that this excess is simply caused by uncertainties un the Solar System decomposition, this may be related also to 
the observation that extremely metal-poor stars with milder or no $r$-process enhancements also have higher first-peak to second-peak $r$-process 
elements ratios than rII stars  (see, e.g \citealt{SiqueiraMelloetal:2014} and this has been known since \citealt{Mcwilliam:1998}).
Both these facts may be understood if $r$-II stars are polluted by the {\em main} $r$-process only, thus hiding the contribution from the {\em weak}
$r$-process that contributes significantly to the first-peak $r$-process elements in the early Galaxy \citep[see, e.g. ][]{Wanajoetal:2006}.

Among elements of the third $r$-process peak, Pt (Z = 78) and Pb (Z = 82)
are most strikingly under-represented by the scaled-Solar $r$-process,
whereas Os and Ir are in reasonable agreement. We note however that,
according to the update of the Ir partition function proposed by
\citet{Mashonkinaetal:2010}, our Ir abundance could be underestimated by
$0.2$\,dex, which would bring Ir up to the same level as Os, some $0.2$
dex above the Solar $r$-process decomposition. 

Concerning Pb, there is now a consensus that the $s$-process
contribution to Pb computed for the Solar System in the decompositions
of \citet{Arlandinietal:1999} and \citet{Simmereretal:2004} were
underestimated \citep[see, e.g.,][ and references
therein]{Roedereretal:2009}. Of central importance, \citet{Arlandinietal:1999}
did not take into account the so-called {\em strong s-process}
contribution to Pb in their decomposition \citep[a component now
identified as the contribution from metal-poor AGB stars integrated over
the history of Galactic chemical evolution, see, e.g.,
][]{Travaglioetal:2001,Travaglioetal:2004}, which would lead to an
increase in the total $s$-process contribution to Pb from 46\% to 83\% 
\citep{Serminatoetal:2009}, thereby decreasing the remaining Solar $r$-process 
Pb abundance by about 0.5~dex compared to the original
\citet{Arlandinietal:1999} decomposition shown in Fig.~\ref{Fig:compth} (upper panel). 
In addition, because Pb is dominated by the $s$-process in the  Solar
System, any uncertainty in the total $s$-process Pb contribution leads
to a substantial variation of the Solar System $r$-process component of
Pb. This is visible also in the variations among the three Solar System
decompositions shown in Fig.~\ref{Fig:compth}. The most recent
decomposition, \citet{Bisterzoetal:2014}, including all the
contributions of the $s$-process to Pb (reaching 87\%), is indeed now in
agreement with the low upper limit we report here for
\object{CS~29497-004} (see Fig.~\ref{Fig:compth}, lower panel).

Thorium and uranium both lie below the Solar System $r$-process
decomposition, as both of these elements have decayed significantly more
in the old star \object{CS~22497-004} than in younger Solar System
material.

\subsubsection{Abundances in \object{CS~29497-004} compared to other
$r$-II stars with available Th and U measurements}\label{sec:comp_rII}

The overall abundance pattern of \object{CS~29497-004} is also very close to that
observed in previously discovered $r$-II stars with Th and U
measurements, the best studied of which are \object{CS~22892-052}
\citep{Snedenetal:1996,Snedenetal:2000,Snedenetal:2003,Snedenetal:2009}, 
\object{CS~31082-001} \citep{Hilletal:2002,Snedenetal:2009,Barbuyetal:2011,SiqueiraMelloetal:2013}, and
\object{HE~1523-0901} \citep{Frebeletal:2007,Snedenetal:2009}. Among the significant
differences that are found, we highlight three:

{\em Uranium and thorium} are lower for \object{CS~29497-004} than
for \object{CS~31082-001},
which displays an anomaly of the heaviest $r$-process nuclei, also known
as the {\em actinide boost} \citep{Hilletal:2002}. In fact,
\cite{Hayeketal:2009} suggested that some 30 to 50\% of all $r$-II stars exhibit
this boost.  \citet{Mashonkinaetal:2014b} refined this estimate to
6 of 18 stars (i.e., 33\%). \object{CS~29497-004} does not belong
to this class, but rather exhibits a ratio of Th to second $r$-process
peak elements compatible with that of \object{CS~22892-052}, which is
also commensurate with a standard $r$-process and the decay of Th over the
age of the Universe.

{\em Platinum} is significantly lower for \object{CS-29497-004} than for
\object{CS~22892-0052}. We note that the Pt abundances for these two
stars were determined from different lines (see Sec.~\ref{detection 3rd
peak}); this may be the reason for the discrepancy. Additionally, our Pt
measurement has a rather large error bar, owing to a blend with an OH
molecular line. Detections of Pt for other $r$-II stars would be most
welcome to investigate the abundance of this third-peak element, and
to validate the use of the 3301\,{\AA} line (conveniently located in the
near-UV part of the spectrum accessible from the ground) to measure its
abundance.

{\em Lead} is a particularly important element, as $\rm ^{208}Pb$ is the
endpoint of many decay chains from heavier unstable products of the
$r$-process, including Th and U. However, in the actinide-boost
star \object{CS~31082-001}, \citet{Plezetal:2004} measured a Pb
abundance that is {\em lower} by some 0.4~dex than both the scaled-Solar
$r$-process fraction of Pb \citep[e.g., $\rm \log\epsilon(Pb/Eu)
=+0.62$ in ][]{Bisterzoetal:2014}, or the $r$-process model predictions:
the theoretical waiting-point approximation $r-$process model of
\citep{Kratzetal:2007} predicts an initial production of $\rm
\log\epsilon(Pb/Eu)_{0}=+0.61$, which, after about 13 Gyrs of decay
of radioactive nuclei, adds up to $\rm \log\epsilon(Pb/Eu)=+0.70$
\citep[see discussion in][]{Roedereretal:2009}. The upper limit for Pb
derived by \citet{Frebeletal:2007} in \object{HE~1523-0901} also implied
a ratio of Pb relative to second- and third-peak $r$-process elements that
was significantly lower than the $r$-process predictions, and
commensurate with the measurement for \object{CS~31082-001}
\citep{Plezetal:2004}. On the other hand, the upper limit for Pb
derived for \object{CS~22892-052} \citep{Roedereretal:2009} is fully
compatible with a scaled-Solar $r$-process. At these low metallicities
($\rm [Fe/H] <-2.5$), these three stars were the only
$r-$process-element dominated stars for which a measurement or
significant upper limit of Pb had been obtained \citep[see the large
sample and compilation by][]{Roedereretal:2010a}.

More recently, \citet{Mashonkinaetal:2012} computed NLTE Pb abundances
for cool metal-poor stars, and showed that the NLTE corrections are
sizable for the \ion{Pb}{I} line used in metal-poor giants, such as the
above $r$-II stars, unlike NLTE corrections for the second-peak $r-$process
element \ion{Eu}{II}, which experiences little departure from NLTE (to be
expected for most elements measured from ionized species). In
particular, \citet{Mashonkinaetal:2012} revised the Pb abundances for
\object{CS~31082-001} to $\rm \log\epsilon(Pb/Eu)_{NLTE} = 0.67$, and
reported an actual detection of Pb for \object{HE~1523-0901} of $\rm
\log\epsilon(Pb/Eu)_{NLTE} = 0.68$ (to be compared to $\rm
\log\epsilon(Pb/Eu)_{LTE} = 0.28$). These new NLTE values now compare
remarkably well with the theoretical waiting-point approximation
$r-$process model for a 13 Gyr old star \citep{Kratzetal:2007,
Roedereretal:2009}, $\rm \log\epsilon(Pb/Eu)=+0.60$, or the Solar
$r-$process decomposition of \citet{Bisterzoetal:2014}, $\rm
\log\epsilon(Pb/Eu)=+0.62$.

The tentative detection of Pb presented here allows us to add
\object{CS~29497-004} to this picture, albeit with a rather large
uncertainty. The most probable LTE Pb abundance for
\object{CS~29497-004} yields $\rm \log\epsilon(Pb/Eu)_{LTE}=+0.41 \pm 0.3$
(while the conservative upper limit is $<+0.91$), to which we apply a
NLTE correction of +0.55 based on \citet{Mashonkinaetal:2012} (their
Table~1, assuming the NLTE correction for their $r-$process enriched
star with $T_{\mathrm{eff}}$=5000K, $\log g$=1.5 and [Fe/H]=$-3.0$),
leading to $\rm \log\epsilon(Pb/Eu)_{NLTE}=+0.96 \pm 0.3$. This is
consistent, within errorbars, with the Pb/Eu ratios in the two other
$r$-II stars where Pb has been detected, \object{CS~31082-001} and
\object{HE~1523-0901}.

Obtaining precise Pb measurements for additional $r$-II stars (with and
without actinide boosts) would clearly be required to better understand the
nature of the $r$-process. One would indeed expect that the actinide
excess in actinide-boost stars should {\em increase} its lead content
with respect to ``normal'' $r-$II stars, owing to the enhanced
contribution of decaying radioactive actinide nuclei to the Pb abundance
over the lifetime of the star. % It is indeed quite striking that an
%$r$-II star with an actinide boost (\object{CS~31082-001}) should
%exhibit the lowest Pb/Eu ratio, since it's actinide excess should, on
%the contrary, {\it increase} its lead content. 

Interestingly, at higher metallicities ($\rm [Fe/H]\sim -2.2$ to
$-1.4$), the reanalysis of the \citet{Roedereretal:2010a} sample in NLTE
by \citet{Mashonkinaetal:2012} found that, over that entire
metallicity range, the Pb/Eu ratio is significantly higher than for the
$r$-II stars discussed above (\object{CS~31082-001},
\object{HE~1523-0901}, and \object{CS~29497-004}), and displays a steady increase (albeit with
some dispersion) towards the Solar Pb/Eu, which is reached around $\rm
[Fe/H]\sim-1.4$. The Pb/Eu ratio in $r$-II stars is assumed to be the
$r-$process yield value, and the steady increase at higher metallicities
is taken as due to early $s-$process enrichment. In this context,
obtaining more Pb/Eu measurements in low-metallicity stars ($r$-II,
$r$-I, and $r$-normal stars) would lead to improved understanding of the
earliest phases of $s$-process enrichment. In particular, do $r$-normal
stars with metallicities below [Fe/H] = $-2.5$ also show a pure $r-$process Pb/Eu
ratio?  Although this would surely be a challenging measurement, it
would also provide significant constraints.

\subsection{Age determinations }

We have determined nucleo-chronometric ages for \object{CS~29497-004}, using
various combinations of chronometers involving Th and U and production
ratios (hereafter PRs) from two 
%models for the r-process:  (1) The 
%high-entropy wind (HEW) models of \citet{Farouqietal:2005}, as reported in
%\citet{Hayeketal:2009}, and using the Th and U theoretical 
%predictions reported in \citet{Roedereretal:2009}, and (2) The
%waiting-point assumption models of \citet{Kratzetal:2007}. 
parameterized nucleosynthesis models developed 
to reproduce the full distribution of the Solar System $r$-process ``residuals" ($\rm N_{r,\odot} = N_{\odot} - N_{s,\odot}$): 
(i) the historical, largely site-independent ``waiting-point" (WP) approach \citep[see, e.g. ][]{B2FH:1957,Kratzetal:1993, Kratzetal:2007}, and (ii) the more recent, site-specific fully dynamical network calculations of the 
high-entropy-wind (HEW) of core-collapse supernovae  \citep[see, e.g. ][]{Woosleyetal:1994, Freiburghausetal:1999, Farouqietal:2010}.

\begin{table*}[htbp]
  \centering
\caption{\label{Tab:AgesCS29497} Age determinations for
    \object{CS~29497-004}, using initial production ratios from: 
(1) $\log(\mathrm{PR})_{\rm HEW}$: HEW predictions (this paper) for $\rm Y_{e}=0.447^{+0.002}_{-0.002}$ based on \citep{Farouqietal:2010}, (2) $\log(\mathrm{PR}_{\rm WP1})$ and (3) $\log(\mathrm{PR}_{\rm WP2})$: waiting-point assumption model predictions \citep[from ][ their Table 2, columns 6 and 3 resp]{Kratzetal:2007}.  Errors on ages are evaluated in two different ways: the $\rm Y_{e}$ variations on PRs are propagated into Age$_{\rm HEW}$ as super- and sub-scripts, while 
ErrAge$_{\mathrm{obs}}$ reports the observational error propagated from the observational uncertainties on the abundance ratio.
For mean ages, the dispersion around the mean is also given in the corresponding age column. 
}
{\renewcommand{\arraystretch}{1.2}
\begin{tabular}{lrcrrrrrrrr}\hline\hline
    X/Y &  \multicolumn{1}{c}{$\log\epsilon(\mathrm{X/Y})_{\mathrm{obs}}$} &
    \multicolumn{1}{c}{ErrAge$_{\mathrm{obs}}$}&
    \multicolumn{1}{c}{$\log(\mathrm{PR})_{\rm HEW}$} &  \multicolumn{1}{c}{Age$_{\rm HEW}$} &
    \multicolumn{1}{c}{$\log(\mathrm{PR}_{\rm WP1})$} &  \multicolumn{1}{c}{Age$_{\rm WP1}$} &
    \multicolumn{1}{c}{$\log(\mathrm{PR}_{\rm WP2})$} &  \multicolumn{1}{c}{Age$_{\rm WP2}$} \\
        &  &   &   &
  (1)   &   &
  (2)   &  &
  (3)   &   \\
        &  &   \multicolumn{1}{c}{[Gyr]} &
   &  \multicolumn{1}{c}{[Gyr]} &
   &  \multicolumn{1}{c}{[Gyr]} &
   &  \multicolumn{1}{c}{[Gyr]} \\
\hline\hline
Th/Ba   & $-1.51 \pm 0.12$ & 5.7& $-1.058^{0.103}_{-0.091}$ &  $21.11^{ 4.81}_{-4.25}$&         &        &         &        \\
Th/La   & $-0.78 \pm 0.08$ & 3.9& $-0.362^{0.105}_{-0.093}$ &  $19.52^{ 4.90}_{-4.34}$&         &        &         &        \\
Th/Ce   & $-0.97 \pm 0.06$ & 3.0& $-0.724^{0.100}_{-0.091}$ &  $11.49^{ 4.67}_{-4.25}$&         &        &         &        \\
Th/Pr   & $-0.49 \pm 0.06$ & 3.0& $-0.313^{0.099}_{-0.092}$ &  $ 8.27^{ 4.60}_{-4.27}$&         &        &         &        \\   %errors were taken to be the mean of Ce and Nd
Th/Nd   & $-1.14 \pm 0.07$ & 3.0& $-0.928^{0.097}_{-0.092}$ &  $ 9.90^{ 4.53}_{-4.30}$&         &        &         &        \\
Th/Sm   & $-0.86 \pm 0.06$ & 3.0& $-0.796^{0.100}_{-0.090}$ &  $ 2.99^{ 4.67}_{-4.20}$&         &        &         &        \\ %new gf Lawler 2006
%Th/Sm   & $-0.90 \pm 0.07$ & 3.2& $-0.796^{0.100}_{-0.090}$ &  $ 4.86^{ 4.67}_{-4.20}$&         &        &         &        \\
Th/Eu   & $-0.50 \pm 0.06$ & 3.0& $-0.240^{0.104}_{-0.092}$ &  $12.14^{ 4.86}_{-4.30}$& $-0.375$&   5.8  & $-0.276$& 10.5   \\
Th/Gd   & $-0.85 \pm 0.07$ & 3.2& $-0.569^{0.105}_{-0.093}$ &  $13.12^{ 4.90}_{-4.34}$&         &        &         &        \\
Th/Dy   & $-1.08 \pm 0.07$ & 3.1& $-0.827^{0.105}_{-0.094}$ &  $11.82^{ 4.90}_{-4.39}$&         &        &         &        \\
Th/Ho   & $-0.40 \pm 0.17$ & 8.1& $-0.071^{0.109}_{-0.046}$ &  $15.36^{ 5.09}_{-2.15}$&         &        &         &        \\
Th/Er   & $-0.87 \pm 0.11$ & 4.9 & $-0.592^{0.107}_{-0.096}$ &  $12.98^{ 5.00}_{-4.48}$&         &        &         &        \\ %new gf Lawler 2008
%Th/Er   & $-0.93 \pm 0.11$ & 5.2& $-0.592^{0.107}_{-0.096}$ &  $15.78^{ 5.00}_{-4.48}$&         &        &         &        \\
Th/Tm   & $ 0.02 \pm 0.08$ & 3.9& $+0.155^{0.106}_{-0.095}$ &  $ 6.31^{ 4.95}_{ 4.44}$&         &        &         &        \\
Th/Hf   & $-0.52 \pm 0.07$ & 3.0& $-0.036^{0.107}_{-0.093}$ &  $22.60^{ 5.00}_{-4.34}$& $-0.335$&   8.6  & $-0.063$& 21.3   \\
\hline			        
%\multicolumn{2}{l}{Th/REE}  weighted mean (rms)     & 3.5&    & $12.42 (\pm 5.15)$     &         &        &         &        \\
\multicolumn{2}{l}{Th/REE}  weighted mean (rms)     & 3.5&    & $11.99 (\pm 5.46)$     &         &        &         &        \\ % updated Sm,Er
\hline			        
Th/Os   & $-1.57 \pm 0.12$ & 5.7& $-0.917^{0.096}_{-0.087}$ &  $30.50^{ 4.48}_{-4.06}$& $-1.125$&   20.8 & $-1.009$&  26.2  \\
Th/Ir   & $-1.39 \pm 0.09$ & 4.1& $-0.839^{0.097}_{-0.089}$ &  $25.73^{ 4.53}_{-4.16}$& $-1.153$&   11.1 & $-1.022$&  17.2  \\
Th/Pt   & $-1.16 \pm 0.22$ &10.1& $-1.112^{0.069}_{-0.065}$ &  $2.24^{ 3.22}_{-3.04}$ & $-1.444$& $-13.3$& $-1.585$& $-19.9$\\
\hline			        
\multicolumn{2}{l}{Th/(Os,Ir)} weighted mean (rms)     & 4.7&      & $27.35 (\pm 2.26)$  && $ 14.4 (\pm 4.6)$& & $20.2 (\pm 4.3)$  \\
\multicolumn{2}{l}{Th/(Os,Ir,Pt)}  weighted mean (rms)& 5.4&       & $24.92 (\pm 7.74)$  &&                  & &                   \\
\hline			        
Th/U    & $-1.04 \pm 0.30$ & 6.6& $0.283^{-0.011}_{+0.013}$ &  $16.50^{-0.24}_{ 0.28}$& $+0.195$&  18.4  & $+0.192$& 18.5   \\
\hline			        
U/Ba 	& $-2.55 \pm 0.32$ & 4.8& $-1.341^{0.114}_{-0.104}$ &  $17.94^{ 1.69}_{-1.54}$&         &        &         &        \\
U/La   	& $-1.82 \pm 0.31$ & 4.6& $-0.645^{0.116}_{-0.106}$ &  $17.44^{ 1.72}_{-1.57}$&         &        &         &        \\
U/Ce   	& $-2.01 \pm 0.30$ & 4.5& $-1.007^{0.111}_{-0.104}$ &  $14.88^{ 1.65}_{-1.54}$&         &        &         &        \\
U/Pr    & $-1.53 \pm 0.30$ & 4.5& $-0.596^{0.110}_{-0.105}$ &  $13.86^{ 1.63}_{-1.55}$&         &        &         &        \\ %errors were taken to be the mean of Ce and Nd
U/Nd   	& $-2.18 \pm 0.30$ & 4.5& $-1.211^{0.108}_{-0.105}$ &  $14.38^{ 1.60}_{-1.56}$&         &        &         &        \\
U/Sm   	& $-1.90 \pm 0.30$ & 4.5& $-1.079^{0.111}_{-0.103}$ &  $12.18^{ 1.65}_{-1.53}$&         &        &         &        \\ %new gf Lawler 2006
%U/Sm   	& $-1.94 \pm 0.30$ & 4.5& $-1.079^{0.111}_{-0.103}$ &  $12.78^{ 1.65}_{-1.53}$&         &        &         &        \\
U/Eu   	& $-1.54 \pm 0.30$ & 4.5& $-0.523^{0.115}_{-0.105}$ &  $15.09^{ 1.71}_{-1.59}$&$-0.570$ &  14.4  & $-0.468$& 15.9   \\
U/Gd   	& $-1.89 \pm 0.30$ & 4.5& $-0.852^{0.116}_{-0.106}$ &  $15.40^{ 1.72}_{-1.57}$&         &        &         &        \\
U/Dy   	& $-2.12 \pm 0.30$ & 4.5& $-1.110^{0.116}_{-0.107}$ &  $14.99^{ 1.72}_{-1.59}$&         &        &         &        \\
U/Ho   	& $-1.44 \pm 0.34$ & 5.1& $-0.354^{0.120}_{-0.059}$ &  $16.12^{ 1.78}_{-0.88}$&          &        &         &        \\
U/Er   	& $-1.91 \pm 0.31$ & 4.6& $-0.875^{0.118}_{-0.109}$ &  $15.36^{ 1.75}_{-1.62}$&         &        &         &        \\ % new gf Lawler 2008
%U/Er   	& $-1.97 \pm 0.32$ & 4.7& $-0.875^{0.118}_{-0.109}$ &  $16.25^{ 1.75}_{-1.62}$&         &        &         &        \\
U/Tm   	& $-1.02 \pm 0.31$ & 4.6& $-0.128^{0.117}_{-0.108}$ &  $13.24^{ 1.74}_{-1.60}$&         &        &         &        \\
U/Hf   	& $-1.56 \pm 0.30$ & 4.5& $-0.319^{0.118}_{-0.106}$ &  $18.42^{ 1.75}_{-1.57}$&$-0.530$ &  15.3  & $-0.356$& 19.4   \\
\hline			        
%\multicolumn{2}{l}{U/REE} weighted mean (rms)              & 4.6&   & $15.41 (\pm 1.68)$      &         &        &         &        \\
\multicolumn{2}{l}{U/REE} weighted mean (rms)              & 4.6&   & $15.29 (\pm 1.74)$      &         &        &         &        \\ %updated Sm,Er
\hline			        
U/Os   	& $-2.61 \pm 0.32$ & 4.7& $-1.200^{0.107}_{-0.100}$ &  $20.92^{ 1.59}_{-1.48}$&$-1.320$ &  19.1  & $-1.201$& 20.9   \\
U/Ir    & $-2.43 \pm 0.31$ & 4.6& $-1.122^{0.108}_{-0.102}$ &  $19.41^{ 1.60}_{-1.51}$&$-1.348$ &  16.1  & $-1.215$& 18.0   \\
U/Pt    & $-2.20 \pm 0.37$ & 5.4& $-1.395^{0.080}_{-0.078}$ &  $11.90^{ 1.19}_{-1.19}$&$-1.639$ &   8.3  & $-1.777$&  6.3   \\ % as given by Kratz (errors made up from "wrong" Th/Pb yields....)
\hline			        
\multicolumn{2}{l}{U/(Os,Ir)} weighted mean (rms)             & 4.6&     & $20.14 (\pm 0.75)$   &&  $ 15.1 (\pm 4.3)$&& $19.4 (\pm 1.4)$ \\
\multicolumn{2}{l}{U/(Os,Ir,Pt}  weighted mean (rms)         & 4.9&     & $17.94 (\pm 3.69)$   &&  $ 17.5 (\pm 1.5)$&& $15.9 (\pm 6.0)$ \\
 \hline
 \end{tabular}
 }	
\end{table*}

%{\it $r$-process production ratios (PRs):}
\subsubsection{$r$-process production ratios (PRs)} \label{sec_PRs}
The WP model used here is based on the historical assumption of an initial Fe seed, and furthermore a 
combination of $\rm (n,\gamma) - (\gamma,n)$ and $\beta$-flow equilibria at high neutron densities ($\rm 10^{20} < n_{n} < 10^{28}$) 
at a fixed temperature ($\rm T_{9} = 1.35$), and an instantaneous freezeout. As discussed in detail in \citet{Kratzetal:1993}, 
in principle a superposition of only three correlated components of neutron densites ($\rm n_{n}$) and nucleosynthesis 
time scales ($\tau_{r}$) to independently fit the three $\rm N_{r,\odot}$ abundance peaks at $\rm A \sim 80$, 130, and 195 was 
required to reproduce the full abundance pattern in nature. An updated version of this model (in terms of a 
finer $\rm n_{n} - \tau_{r}$ grid and an improved nuclear-physics input) from \citet{Kratzetal:2007} is used for the age 
determinations presented in Table~\ref{Tab:AgesCS29497}. 

A more recent, still commonly used $r$-process model applied here, involves the so-called neutrino or high-entropy 
wind (HEW) from Type II (core-collapse) supernovae \citep[ccSN; see, e.g., ][]{Woosleyetal:1994}. In the calculations
used for the present chronometric age determinations, we follow the description of adiabatically expanding 
mass zones at the outermost neutron star layers, as initially utilized in \citet{Freiburghausetal:1999}. After several
updates of both the detailed dynamical simulations, as well as the theoretical and experimental nuclear-data 
input, we use here the version explained in great detail in \citet{Farouqietal:2010}. 
After a charged-particle freezeout leading to a neutron-rich {\em seed} composition beyond the historically assumed 
Fe peak and even beyond the $\rm N = 50$ magic shell, the expanding mass zones have different initial entropies (S), 
so that in the HEW model the overall ejected $r$-process matter represents a superposition of entropies. The     
ratio of free neutrons to  {\em seed} nuclei ($\rm Y_{n}/Y_{seed}$) is correlated to the three main parameters of the HEW, 
i.e. the electron abundance ($\rm Y_{e}$ = Z/A, which defines the {\em neutron richness} of the $r$-ejecta), the entropy S 
and the expansion velocity of the ejecta ($\rm Y_{exp}$); and can be expressed by a simple {\em r-process strength}
formula $\rm Y_n/Y_{seed} \sim V_{exp}(S/Y_e)^{3}$. Within this parameter space, it became immediately evident that 
the ccSN - HEW model predicts two clearly different primary, rapid nucleosynthesis modes with three different 
components: at low entropies (S $< 50$) a neutron-rich charged-particle component without free neutrons, for 
somewhat higher entropies (50 $< S < $150) a {\em weak} neutron-capture r-process, and for high entropies 
(S $> 150$) a robust Solar System-like {\em main}  $r$-process, which we need here for our age determinations \citep[for 
details, see ][]{Kratzetal:2008, Farouqietal:2009a, Farouqietal:2009b, Farouqietal:2010}.  
Initially, as in the case of the WP approach, purposively chosen to closely reproduce the full $\rm N_{r,\odot}$ abundance pattern, 
a parameter combination of $\rm Y_e = 0.45, V_{exp} = 7500$\,km/s and an entropy range of $\rm 10 < S < 280$ was chosen 
to compare our HEW predictions to recent astronomical observations \citep[see, e.g.][]{Farouqietal:2009a, Hayeketal:2009, Farouqietal:2010, Kratzetal:2014}. 
This approach seemed to be justified for quite some time in 
the past, because the striking similarity between the $N_{r,\odot}$ abundances above Ba \citep[in fact already above Te, 
see ][]{Roedereretal:2012_Tellurium} with the pattern observed in the EMP $r$-II star \object{CS~22892-052} \citep{Snedenetal:1996, Snedenetal:2003}, 
has led to the conclusion that every astrophysical event will produce such a ``unique" Solar-like pattern. However, 
more recently the observations of several so-called actinide-boost stars (such as \object{CS 31082-001} \citep{Hilletal:2002, Barbuyetal:2011, SiqueiraMelloetal:2013}, 
or \object{HE~2252-4225} \citep{Mashonkinaetal:2014b}),  and the $r$-process ubiquity study of \citet{Roedereretal:2010a}, have 
questioned this assumption. 
Therefore, motivated by these observations, and our failure to obtain a consistent mean age for the Th/X and U/X
chronometer pairs over the entire Solar-like rare earth elements and third-peak region, we have extended our HEW age determinations from the 
initial constant $\rm Y_e - V_{exp} - S$ parameter assumption to variations of the $Y_e$ parameter. With this, we are changing the neutron 
richness of the $r$-process ejecta, to represent the diversity of observed $r$-process. And indeed, with optimized values of $Y_{e} < 0.45$ we 
have been able to obtain more consistent mean chronometric age. For example, for $r$-II stars showing an actinide boost (e.g., \object{CS 31082-001}), we
can thus avoid the unphysical negative age for Th/Eu that was derived when using PRs reproducing the Solar System $r$-process.
In the present case of \object{CS 29497-004}, we have been able to obtain consistent mean Th/X, U/X and 
Th/U ages for a slightly lower electron fraction $\rm Y_e \sim 0.447 (\pm0.002)$, i.e., for HEW ejecta that are more 
neutron rich than required to fit the standard $\rm N_{r,\odot}$ ones. 

The resulting ages are listed in
Table~\ref{Tab:AgesCS29497}, where the HEW model PRs and ages are given in columns (1) and the WP approach are reported in columns (2) and (3) as in \citet{Kratzetal:2007}. The age uncertainties (ErrAge$_{\mathrm{obs}}$)
listed in this table take into account the observational errors -- 
that is, the uncertainties in the determination of the abundance ratios, including
measurement errors and (random) errors of stellar parameters ($T_{\mathrm{\tiny eff}}$ and
$\log g$). Uncertainties associated to the assumed set of PRs are explored both by comparing HEW and WP models, and through the exploration of different  $\rm Y_{e}$ within the HEW model (reported as super- and sub-scripts for lower and upper uncertainties on the PR$_{\rm HEW}$ and corresponding Age$_{\rm HEW}$ columns). Comparing age uncertainties arising from PRs to ErrAge$_{\mathrm{obs}}$ in this table, it can be appreciated that, while for Th based ratios, the two sources have comparable amplitudes, the observational uncertainty always dominates chronometers involving U.

%{\it WP approximation model ages:}
\subsubsection{WP approximation model ages}
Table~\ref{Tab:AgesCS29497} also presents ages derived using PRs from
\citep{Kratzetal:2007} based on $r$-process calculations using Fe seeds.
In column (3) we report the theoretical production ratio results that yield the best overall fit to
the Solar System $r$-process abundance data for masses $\mathrm{A} > 83$, while in column
(2) we report the ratios of theoretical production of Th and U to
the present-day {\em observed} Solar System $r$-process elemental abundances 
\citep[see ][ for a detailed description of column 6 of their Table 2]{Kratzetal:2007} . 
These ages display the same basic features as the
HEW PRs, namely older ages when U is used rather than Th, in combination
with any of the rare earths, but still commensurate within
uncertainties. We further note that PRs computed with respect to
present-day Solar System elemental abundances (col. 2) yield ages that
are, on average, lower by $7.2$\,Gyr for Th-based and $2.3$\,Gyrs for U-based
chronometers than those using purely theoretical PRs (col 3). With these
PRs computed with respect to
present-day Solar System elemental abundances, 
the weighted mean age for \object{CS~29497-004} is $12.2$\,Gyr, with
a dispersion around the mean of $\sigma = 4.9$\,Gyr, and a formal
(observational) error of $3.6$\,Gyr.

%{\it HEW-model ages:}
\subsubsection{HEW-model ages}
The weighted average of the age determinations for \object{CS~29497-004}
using the ratios of Th to 12 rare-earth species (Ba, La, Ce, Pr, Nd, Sm,
Eu, Gd, Dy, Ho, Er, Tm) and the HEW PRs with $\rm Y_e=0.447$ is $12.0$\,Gyr, with a weighted dispersion
around the mean of $\sigma = 5.4$\,Gyr, and a formal (observational)
error of $3.5$\,Gyr. The weighted average age using the ratios of U to
the rare-earth ratios is somewhat higher: $15.3$\,Gyr, with a
significantly lower weighted dispersion around the mean of $\sigma =
1.7$\,Gyr, and a formal (observational) error of $4.6$\,Gyr. The observational age
uncertainty is in both cases due mostly to the observational uncertainty
on the Th (or U) abundance itself, and for U-based chronometers, observational uncertainties 
largely exceed uncertainties due to PRs. These two mean ages are therefore
both fully compatible with each other, and with the accepted age of the
Universe of $13.772 \pm 0.059$\,Gyr \citep{Bennettetal:2013}. The Th/U chronometer pair alone yields an age of
$16.5$\,Gyr with an associated observational uncertainty of $6.6$\,Gyr,
again fully compatible with the age of the Universe. The combined age of
\object{CS~29497-004} from all these chronometer pairs is $13.7$\,Gyr,
with a formal (observational) error of $4.4$\,Gyr. 
The low scatter
around the mean age from the ratios of U to the rare earths indicates
that the production ratios for these elements are well-predicted by the
HEW model. 
The chronometer pairs using the heaviest detected second-peak
element Hf and the third-peak element Os provide significantly higher
ages, hinting at possible Hf and Os over-production in the HEW model.

As outlined in section~\ref{sec_PRs}, the theoretical predictions of
the PRs of the relevant chronometer pairs also suffer uncertainties linked to the many flavours of $r$-process that 
can arise in astrophysical sites, for example linked to the neutron richness of the ejecta. 
To quantify this effect, we tabulated in Table~\ref{Tab:AgesCS29497} PRs arising from 
the HEW model with different $\rm Y_e$ (given as lower and upper uncertainties to PRs) 
for each chronometer pair (Th to each of the Rare Earth Elements -hereafter REE-, U to REE elements, U/Th, etc.). 
Fig.~\ref{Fig:age_ye} shows how ages react as a function of $\rm Y_e$. Because the 
slope of the age vs. $\rm Y_e$ is different for different chronometer pairs (e.g., Th/REE, 
U/REE, and U/Th), it is in principle be possible to determine the best fitting $\rm Y_e$ for a given 
star. For \object{CS~29497-004}, all $\rm Y_e$ explored here are allowed, and both the chosen $\rm Y_e=0.447$ or $\rm Y_e=0.445$ allow
an excellent agreement for the age of different chronometer pairs.

\begin{figure}[htbp]
  \centering
  \includegraphics[width=8.5cm]{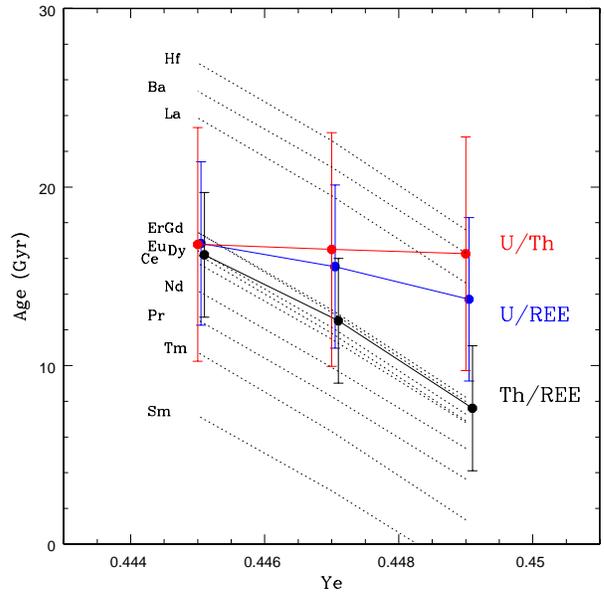}
  \caption{Ages for \object{CS~29497-004} as a function of the assumed neutron richness ($\rm Y_e$) in the HEW model: weighted mean of Th/X ages for 12  rare-earth elements
(black circles and full line);  weighted mean of U/X ages for 12 rare-earth elements (blue circles and full line); U/Th age (red circles and full line). The black dotted line depict Th/X ages for the individual elements. Weights are used to take into account the observational error on abundance ratios.}\label{Fig:age_ye} 
\end{figure}

Comparing PRs and ages obtained through various methods, as is done
in Table~\ref{Tab:AgesCS29497} by comparing the HEW and waiting-point
approximation PRs, also allows to gain insight on the robustness of theoretical predictions. 
 For the ratios for which all three PRs are
available, the variation among PRs is lower than the observational
uncertainty for all pairs involving U (Th/U, U/Eu, U/Os, U/Ir, U/Pt) except U/Hf, while the
variation among the PR sources is larger than the observational
uncertainty for pairs involving the Th chronometer alone (Th/Eu, Th/Hf, Th/Os, Th/Ir, Th/Pt). 
Most remarkable is the agreement
obtained for Th/U, confirming that this ratio is a robust chronometer
even in the face of PR uncertainties
\citep[e.g.,][]{Goriely/Arnould:2001,Hilletal:2002,Wanajoetal:2002}. 
Fig.~\ref{Fig:age_ye} 
illustrates again the robustness of the Th/U chronometer pair, as the production ratios of these two elements move in 
lockstep when physical conditions change. 

%Furthermore, \citet{Frebeletal:2007} collected PRs from the literature
%to investigate the range of published values and the effect that the
%choice of different PRs has on nucleo-chronometric age determinations.
%The result is that ages determined with chronometers involving Th can
%vary by up to 3.8\,Gyr; the corresponding variation for chronometers
%involving U is 1.8\,Gyr. Most of the PRs from HEW (column 1) and
%\citet{Kratzetal:2007} (column 2) that we adopted here lie within, or
%very close to, the range of values collected by \citet{Frebeletal:2007}.
%The lone exceptions are the pairs involving U and Th relative to Os,
%where the HEW PRs lead to ages higher by 2.8 and 10.2\,Gyr, respectively,
%as compared to the age that would result from the PR of
%\citet{Schatzetal:2002}. 

%Th/Hf, U/Hf have recently been argued to provide particularly robust
%chronometer pairs \citep{Kratzetal:2007}, but 

Finally, we note that the very low Pt abundance in \object{CS~29497-004}
leads to exceedingly low (and even unphysical) ages when used as a
reference element for both of the Th or U chronometer pairs, which we suspect
is likely due to a systematic offset of our Pt measurement.

\section{Discussion and conclusions}\label{Sect:DiscussionConclusions}

The discovery of $r$-II stars has proceeded slowly since their first
recognition over twenty years ago. There now exist a total of about 25
$r$-II stars that have been identified among stars of the Galactic halo
population, but only three (now including \object{CS~29497-004}) have
confident detections of U. The majority of the remaining $r$-II stars
have measurements of Th available as well. One additional halo $r$-II star
with detected U (and Th) has been found from high-resolution
spectroscopic follow-up of VMP/EMP stars from the RAVE survey
\citep{Steinmetzetal:2006}, as reported by Placco et al. (in prep.). Although
it is clearly desirable to dramatically increase the numbers of known
$r$-II stars (efforts to accomplish this goal are presently underway),
the extant sample already allows a number of broad conclusions to be
reached.

\begin{itemize}

\item{\bf $r$-II stars are found in all evolutionary stages}:  Until recently, 
all of the recognized $r$-II stars were giants, leaving open the
possibility that there exists some form of peculiarity that could account
for this enhancement specifically in red-giant stellar atmospheres \citep[see, e.g., ][]{Snedenetal:2008}.
The identification of SDSS~J2357-0052 as a
cool main-sequence $r$-II star by \citet{Aokietal:2010} removed much of this
concern. Even more recently, \citet{Roedereretal:2014} identified a total of
nine new highly $r$-process-enhanced stars from the HK survey (a mix of
$r$-I and $r$-II stars) in the subgiant and horizontal-branch stages of
evolution, confirming that this phenomenon occurs in all stages, and
almost certainly arose due to pollution of the natal gas from which the
$r$-II and $r$-I stars formed by progenitor(s) that
produced copious amounts of $r$-process elements.\\    

\item{\bf $r$-II stars that exhibit an actinide boost}: As
shown in paper X of this series \citep{Mashonkinaetal:2014b}, 
on the order of 30\% of $r$-II stars exhibit substantially
higher abundances of Th (and U, where it is observed) relative to
stable $r$-process elements such as Eu (by $\sim 0.3-0.4$ dex)
compared to the majority of $r$-II stars.  This so-called actinide-boost
phenomenon has not yet been accounted for by models for the
progenitors that created the $r$-process, but it remains, in our view, a
strong potential constraint, in particular if a single class of
progenitors are ultimately identified with the origin of $r$-process
elements.  They must be capable of, about one-third of the time, 
producing this {\em obvious} abundance anomaly. Clearly, refinement of 
the observed frequency of actinide-boost stars among highly
$r$-process-element enhanced stars is desired.\\

\item{\bf Contrast in the first-peak $r$-process elements among
$r$-I and $r$-II stars}: \citet{SiqueiraMelloetal:2014} have carried out a
detailed comparison of the first-peak (e.g., Sr, Y, Zr) relative to the second-peak
(e.g., Ba, La, Eu) elements for known $r$-I and $r$-II stars, concluding
that the first-peak elements (generally associated with the weak
$r$-process) are enhanced in the $r$-I stars relative to those observed
in the $r$-II stars. In other words, the weak $r$-process signature 
observed in $r$-normal and $r$-I stars is drowned in the excess of 
main $r$-process that define the $r$-II class. 
Although confirmation of this result awaits
determinations of the abundance patterns for a larger number of
$r$-process-enhanced stars, these authors suggest that this could be
interpreted as evidence that different nucleosynthesis pathways (hence
different progenitor sites) are responsible for the different
sub-classes of $r$-process-enhanced stars.\\

\item{\bf Rejection of binary mass-transfer scenarios associated with 
$r$-process enhancement}: Previously suggested models for the production
of enhancements in $r$-process elements for a small fraction of VMP/EMP
stars considered the possibility that these arose from mass-transfer
events (via wind accretion) associated with a companion star responsible
for their creation \citep[see, e.g., ][and references
therein]{Wanajoetal:2006}. Limited radial-velocity monitoring campaigns
were unable to reject these models for a number of $r$-II stars,
including the prototypical star \object{CS~22892-052} \citep[e.g.,
][]{Preston/Sneden:2001}. The results of a more extensive
radial-velocity monitoring program (based on far more accurate
measurements than were previously available) for $r$-process-enhanced
stars, reported by \citet{Hansenetal:2015}, have demonstrated
convincingly that the chemical peculiarities of the $r$-II and $r$-I
stars cannot be generally caused by binary companions. The binary
fraction of such stars was shown by these authors to be 18$\pm$11\% (3
of 17 stars), completely consistent with the
observed binary fractions of chemically ``normal'' VMP and EMP stars.
Instead, as these authors concluded, the $r$-process enhancement was
imprinted on the natal molecular clouds of these stars by an external
source. 

\end{itemize}

 Our study of the third $r$-II star with detected U,
\object{CS~29497-004}, an object that does not exhibit an actinide
boost, has enabled new estimates of the radioactive-decay age for $r$-II
stars, albeit with larger than desired errors arising due to the
difficulty of obtaining accurate abundances for chronometers involving U
for this relatively faint ($V = 14$) and comparatively warm
($T_{\mathrm{eff}} = 5000$~K) star. We have also obtained a tentative
detection of Pb for this star, and called attention to the important
role this element plays in providing constraints on the nature of the
$r$-process, and the rise of the $s$-process.  We look forward to
the refinement of our understanding expected from the identification of additional
$r$-process-enhanced stars with detected Pb, Th, and U, both with and
without actinide boosts, based on newly initiated surveys of (in particular {\it
bright}; $9 < V < 14$) VMP and EMP stars now underway.

\begin{acknowledgements}
  
We are grateful to the ESO staff at Paranal and Garching for obtaining
the observations and reducing the data, respectively. N.C. expresses his
gratitude to the Observatoire Paris-Meudon for hospitality shown to him
during a 1-month visit during which most of this paper was written. He 
acknowledges support by Sonderforschungsbereich SFB 881 "The Milky
 Way System" (subproject A4) of the German Research Foundation (DFG).
T.C.B. acknowledges partial support for
this work from grants PHY 02-16783, PHY 08-22648; Physics Frontier
Center/Joint Institute for Nuclear Astrophysics (JINA), and PHY 14-30152;
Physics Frontier Center/JINA Center for the Evolution of the Elements
(JINA-CEE), awarded by the US National Science Foundation.
P.S.B. is a Research Fellow of the Royal Swedish Academy of Sciences
supported by a grant from the Knut and Alice Wallenberg Foundation.
%[ANYTHING ELSE?]

\end{acknowledgements}

%________________________________________________________________________________
%
\bibliographystyle{aa}
\bibliography{HES,mphs,ncastro,ncpublications,photometry,vanessa}

\begin{appendix}

\section{Individual radial velocity measurements for  \object{CS~29497-004}}

\begin{table}[htbp]
 \centering
 \caption{Radial velocity data for \object{CS~29497-004}.}
 \label{tab:vrad}
  \begin{tabular}{lll}\hline\hline
   HJD    & $v_{\mathrm{rad}}{\mathrm km/s}$ & $\sigma$\\\hline
   2452214.634255 & 105.023 & 0.502 \\
   2452549.624812 & 104.039 & 0.395 \\
   2452555.600435 & 104.173 & 0.271 \\
   2452582.517666 & 104.697 & 0.350 \\
   2452584.522592 & 104.536 & 0.345 \\
   2452584.630191 & 104.355 & 0.316 \\
   2452584.674641 & 104.514 & 0.324 \\
   2452586.639595 & 104.872 & 0.278 \\
   2452586.684487 & 104.722 & 0.907 \\
   2452607.617861 & 104.120 & 0.253 \\
   2452607.571902 & 104.322 & 0.303 \\
   2452608.601247 & 103.638 & 0.244 \\
   2452608.557655 & 104.046 & 0.142 \\
   2452609.536427 & 104.284 & 0.384 \\
   2452609.581567 & 103.868 & 0.581 \\
   2452615.544251 & 104.413 & 0.401 \\
   2452615.587581 & 104.202 & 0.434 \\
   2452616.537721 & 104.621 & 0.377 \\
   2452616.581430 & 104.581 & 0.376 \\
   2452632.539782 & 104.330 & 0.574 \\
   2452633.540846 & 104.363 & 0.669 \\
   2452634.541717 & 104.735 & 0.313 \\
   2452635.543508 & 104.367 & 0.404 \\
   2452826.830219 & 105.199 & 0.536 \\
   2452826.874671 & 105.211 & 0.269 \\
   2452853.759219 & 105.123 & 0.292 \\
   2452858.869418 & 104.831 & 0.364 \\
   2452859.880800 & 104.476 & 0.351 \\
   2452882.681706 & 104.541 & 0.204 \\
   2454373.642892 & 104.329 & 0.116 \\
   2454705.631179 & 104.873 & 0.065 \\
   2454780.500993 & 105.409 & 0.292 \\
   2454819.320354 & 105.561 & 0.084 \\
   2455175.419215 & 105.010 & 0.063 \\
   2455415.686599 & 104.941 & 0.061 \\
   2455439.606743 & 105.124 & 0.060 \\
   2455796.671082 & 105.434 & 0.077 \\
   2455858.545060 & 105.060 & 0.060 \\
   2456191.589163 & 105.050 & 0.045 \\
   2456530.687786 & 104.829 & 0.054 \\
   2456956.535485 & 104.471 & 0.094 \\
  \end{tabular}
\end{table}

\section{Neutron-capture element linelist}

Table \ref{Tab:LinelistHeavy} reports the heavy elements linelist together with individual abundance
measurements for \object{CS~29497-004}. References for $\rm \log gf$ sources are given in column 5, as follows: 
1: \citet{Snedenetal:2003}; 
2: \citet{Snedenetal:1996}; 
3: \citet{CowleyCorliss:1983}; 
4: \citet{BGZr2:1981};
5: \citet{BGRu1:1984};
6: \citet{HLBNb2:1985}; 
7: \citet{WBMo:1988}; 
8: \citet{SLRu1:1985};
9: \citet{KZNb1Rh1:1982}; 
10: \citet{DLRh1:1985};
11: \citet{CB:1962}; 
12: \citet{JMGAg1:1978}; 
13: \citet{NBSBa:1969}; 
14:\citet{Mcwilliam:1998}; 
15: \citet{Lawleretal:2001a}
16: DREAM database \citep{Palmerietal:2000}; 
17: \citet{Ivarssonetal:2001}; 
18: \citet{Denhartogetal:2003}; 
19: \citet{Lawleretal:2006};
20: \citet{Lawleretal:2001b}; 
21: \citet{Lawleretal:2001c}; 
22: \citet{Lawleretal:2004}; 
23:\citet{Lawleretal:2008}; 
24: \citet{Lawleretal:2007};
25: \citet{Ivarssonetal:2003}; 
26: \citet{NGPb:1969}; 
27: \citet{Nilssonetal:2002b}; 
28: \citet{Nilssonetal:2002a}.
References for hyperfine splitting parameters used are given in column 6, as follows: 
1: \citet{Mcwilliam:1998}; 
2:  \citet{Lawleretal:2001a}; 
3: \citet{Ivarssonetal:2001}; 
4: \citet{Lawleretal:2001b}; 
5:\citet{Lawleretal:2001c}; 
6:\citet{Lawleretal:2004}; 
7: \citet{Martensson-Pendrilletal:1994}.

%\longtab{}{ 
\begin{longtable}{rrrrllcr} 
\caption{\label{Tab:LinelistHeavy} Neutron-capture elements linelist
         and individual abundances in CS~29497-004.}\\
\hline\hline 
Species     & $\lambda$(\AA) & E.P.(eV) & $\rm \log gf$ & Ref$_{\rm \log gf}$ \footnote{
References for $\rm \log gf$ sources: 
1: \citet{Snedenetal:2003}; 
2: \citet{Snedenetal:1996}; 
3: \citet{CowleyCorliss:1983}; 
4: \citet{BGZr2:1981};
5: \citet{BGRu1:1984};
6: \citet{HLBNb2:1985}; 
7: \citet{WBMo:1988}; 
8: \citet{SLRu1:1985};
9: \citet{KZNb1Rh1:1982}; 
10: \citet{DLRh1:1985};
11: \citet{CB:1962}; 
12: \citet{JMGAg1:1978}; 
13: \citet{NBSBa:1969}; 
14:\citet{Mcwilliam:1998}; 
15: \citet{Lawleretal:2001a}
16: DREAM database \citep{Palmerietal:2000}; 
17: \citet{Ivarssonetal:2001}; 
18: \citet{Denhartogetal:2003}; 
19: \citet{Lawleretal:2006};
20: \citet{Lawleretal:2001b}; 
21: \citet{Lawleretal:2001c}; 
22: \citet{Lawleretal:2004}; 
23:\citet{Lawleretal:2008}; 
24: \citet{Lawleretal:2007};
25: \citet{Ivarssonetal:2003}; 
26: \citet{NGPb:1969}; 
27: \citet{Nilssonetal:2002b}; 
28: \citet{Nilssonetal:2002a}
}
& Ref$_{\rm HFS}$\footnote{
References for HFS source: 
1: \citet{Mcwilliam:1998}; 
2:  \citet{Lawleretal:2001a}; 
3: \citet{Ivarssonetal:2001}; 
4: \citet{Lawleretal:2001b}; 
5:\citet{Lawleretal:2001c}; 
6:\citet{Lawleretal:2004}; 
7: \citet{Martensson-Pendrilletal:1994}}
            &    \multicolumn{1}{l}{E.W.(m\AA)} &$\rm \log \epsilon$\\
\hline
\endfirsthead 
\caption{continued.}\\ 
\hline\hline 
Species     & $\lambda$(\AA) & E.P.(eV) & $\rm \log gf$ & Ref$_{\rm \log gf}$ & Ref$_{\rm HFS}$ &
                \multicolumn{1}{l}{E.W.(m\AA)} &$\rm \log \epsilon$\\
\hline 
\endhead 
\hline 
\endfoot 
%  \begin{tabular}{rrrrlcr}\hline\hline
%Species     & $\lambda$(\AA) & E.P.(eV) & $\rm \log gf$ &
%Ref. &
%  \multicolumn{1}{l}{E.W.(m\AA)} &$\rm \log \epsilon$\\\hline
\ion{Sr}{II} & 3464.45 &  3.04 & $ 0.530$ & 1 &   &    syn    & $ 0.67$ \\
\ion{Sr}{II} & 4077.71 &  0.00 & $ 0.170$ & 2 &   &    syn    & $ 0.67$ \\
\ion{Sr}{II} & 4161.79 &  2.94 & $-0.600$ & 2 &   & syn(5.4) & $ 0.87$ \\
\ion{Sr}{II} & 4215.52 &  0.00 & $-0.170$ & 2 &   &    syn    & $ 0.77$ \\
\ion{ Y}{II} & 3242.28 &  0.18 & $ 0.210$ & 1 &   &   65.9    & $-0.08$ \\
\ion{ Y}{II} & 3327.88 &  0.41 & $ 0.130$ & 1 &   &   56.3    & $-0.09$ \\
\ion{ Y}{II} & 3549.01 &  0.13 & $-0.280$ & 1 &   &   52.8    & $-0.17$ \\
\ion{ Y}{II} & 3600.74 &  0.18 & $ 0.280$ & 1 &   &   71.6    & $-0.12$ \\
\ion{ Y}{II} & 3611.04 &  0.13 & $ 0.110$ & 1 &   &   63.6    & $-0.27$ \\
\ion{ Y}{II} & 3747.56 &  0.10 & $-0.910$ & 1 &   &   38.7    & $-0.01$ \\
\ion{ Y}{II} & 3774.33 &  0.13 & $ 0.210$ & 1 &   &   75.8    & $-0.13$ \\
\ion{ Y}{II} & 3788.69 &  0.10 & $-0.070$ & 1 &   &   73.6    & $ 0.04$ \\
\ion{ Y}{II} & 3818.34 &  0.13 & $-0.980$ & 1 &   &   41.8    & $ 0.14$ \\
\ion{ Y}{II} & 3832.90 &  0.18 & $-0.340$ & 1 &   &   56.0    & $-0.11$ \\
\ion{ Y}{II} & 3950.35 &  0.10 & $-0.490$ & 1 &   &   63.9    & $ 0.12$ \\
\ion{ Y}{II} & 4398.01 &  0.13 & $-1.000$ & 1 &   &   39.3    & $ 0.01$ \\
\ion{ Y}{II} & 4883.68 &  1.08 & $ 0.070$ & 2 &   &   35.0    & $-0.10$ \\
\ion{Zr}{II} & 3334.61 &  0.56 & $-0.797$ & 1 &   &   30.6    & $ 0.78$ \\
\ion{Zr}{II} & 3338.41 &  0.96 & $-0.578$ & 1 &   &   11.3    & $ 0.42$ \\
\ion{Zr}{II} & 3410.24 &  0.41 & $-0.323$ & 1 &   &   59.8    & $ 0.91$ \\
\ion{Zr}{II} & 3424.81 &  0.04 & $-1.305$ & 1 &   &   27.2    & $ 0.59$ \\
\ion{Zr}{II} & 3430.51 &  0.47 & $-0.164$ & 1 &   &   48.1    & $ 0.46$ \\
\ion{Zr}{II} & 3438.23 &  0.09 & $ 0.310$ & 3 &   &   88.9    & $ 0.86$ \\
\ion{Zr}{II} & 3457.55 &  0.56 & $-0.530$ & 1 &   &   39.7    & $ 0.71$ \\
\ion{Zr}{II} & 3458.92 &  0.96 & $-0.520$ & 1 &   &   13.0    & $ 0.42$ \\
\ion{Zr}{II} & 3479.03 &  0.53 & $-0.690$ & 1 &   &   26.3    & $ 0.50$ \\
\ion{Zr}{II} & 3479.38 &  0.71 & $ 0.170$ & 1 &   &   51.4    & $ 0.48$ \\
\ion{Zr}{II} & 3499.56 &  0.41 & $-0.810$ & 1 &   &   22.5    & $ 0.38$ \\
\ion{Zr}{II} & 3505.68 &  0.16 & $-0.360$ & 1 &   &   59.2    & $ 0.61$ \\
\ion{Zr}{II} & 3536.93 &  0.36 & $-1.306$ & 1 &   &   13.3    & $ 0.52$ \\
\ion{Zr}{II} & 3551.94 &  0.09 & $-0.310$ & 1 &   &   65.5    & $ 0.67$ \\
\ion{Zr}{II} & 3573.05 &  0.32 & $-1.041$ & 1 &   &   30.7    & $ 0.71$ \\
\ion{Zr}{II} & 3578.21 &  1.21 & $-0.607$ & 1 &   &    9.4    & $ 0.61$ \\
\ion{Zr}{II} & 3630.00 &  0.36 & $-1.110$ & 1 &   &   25.6    & $ 0.69$ \\
\ion{Zr}{II} & 3698.15 &  1.01 & $ 0.094$ & 1 &   &   42.2    & $ 0.56$ \\
\ion{Zr}{II} & 3714.79 &  0.53 & $-0.930$ & 1 &   &   34.2    & $ 0.85$ \\
\ion{Zr}{II} & 3751.61 &  0.97 & $ 0.012$ & 1 &   &   40.1    & $ 0.54$ \\
\ion{Zr}{II} & 3766.80 &  0.41 & $-0.812$ & 1 &   &   40.4    & $ 0.73$ \\
\ion{Zr}{II} & 3836.76 &  0.56 & $-0.060$ & 4 &   &   56.1    & $ 0.50$ \\
\ion{Zr}{II} & 3998.95 &  0.56 & $-0.387$ & 3 &   &   47.1    & $ 0.58$ \\
\ion{Zr}{II} & 4050.32 &  0.71 & $-1.000$ & 1 &   &   17.0    & $ 0.64$ \\
\ion{Zr}{II} & 4090.53 &  0.76 & $-1.009$ & 1 &   &   16.1    & $ 0.67$ \\
\ion{Zr}{II} & 4161.21 &  0.71 & $-0.720$ & 1 &   &   36.3    & $ 0.82$ \\
\ion{Zr}{II} & 4208.98 &  0.71 & $-0.460$ & 1 &   &   38.7    & $ 0.61$ \\
\ion{Nb}{II} & 3215.59 &  0.44 & $-0.190$ & 6 &   & syn(21.0) & $-0.33$ \\
\ion{Mo}{I } & 3864.10 &  1.00 & $-0.010$ & 7 &   &    syn    & $ 0.15$ \\
\ion{Ru}{I } & 3436.74 &  0.15 & $ 0.015$ & 1 &   & syn(11.7) & $ 0.64$ \\
\ion{Ru}{I } & 3498.94 &  0.00 & $ 0.310$ & 8 &   & syn(25.6) & $ 0.62$ \\
\ion{Ru}{I } & 3728.03 &  0.00 & $ 0.270$ & 5 &   &    syn    & $ 0.55$ \\
\ion{Ru}{I } & 3798.90 &  0.15 & $-0.040$ & 5 &   & syn(10.6) & $ 0.56$ \\
\ion{Ru}{I } & 3799.35 &  0.00 & $ 0.020$ & 5 &   & syn(19.5) & $ 0.66$ \\
\ion{Rh}{I } & 3434.89 &  0.00 & $ 0.450$ & 9 &   & syn(11.3) & $-0.27$ \\
\ion{Rh}{I } & 3692.36 &  0.00 & $ 0.174$ & 10&   & syn(12.3) & $-0.10$ \\
\ion{Pd}{I } & 3242.70 &  0.81 & $-0.070$ & 5 &   &    syn    & $ 0.31$ \\
\ion{Pd}{I } & 3404.58 &  0.81 & $ 0.320$ & 5 &   & syn(23.6) & $-0.07$ \\
\ion{Pd}{I } & 3460.74 &  0.81 & $-0.420$ & 5 &   &    syn    & $ 0.11$ \\
\ion{Pd}{I } & 3516.94 &  0.96 & $-0.240$ & 5 &   & syn(10.4) & $ 0.18$ \\
\ion{Pd}{I } & 3634.69 &  0.81 & $ 0.090$ & 11&   &   36.2    & $ 0.44$ \\
\ion{Ag}{I } & 3280.68 &  0.00 & $-0.050$ & 12&   & syn(32.5) & $-0.58$ \\
\ion{Ag}{I } & 3382.89 &  0.00 & $-0.377$ & 12&   & syn(19.2) & $-0.38$ \\
%\ion{Cd}{I } & 3261.05 &  0.00 & $-2.470$ & &   &    0.0    & $ 0.00$ \\
%\ion{Sn}{I } & 3262.33 &  1.07 & $ 0.110$ & &   &    0.0    & $ 0.00$ \\
%\ion{Sn}{I } & 3801.01 &  1.07 & $-0.620$ & &   &    0.0    & $ 0.00$ \\
\ion{Ba}{II} & 3891.78 &  2.51 & $ 0.280$ & 13 &   &    syn    & $ 0.22$ \\
\ion{Ba}{II} & 4130.65 &  2.72 & $ 0.560$ & 14 & 1&    syn    & $ 0.32$ \\
\ion{Ba}{II} & 4166.00 &  2.72 & $-0.420$ & 13 &   &    syn    & $ 0.52$ \\
\ion{Ba}{II} & 4554.03 &  0.00 & $ 0.170$ & 14 &   1&    syn    & $ 0.52$ \\
\ion{Ba}{II} & 4934.08 &  0.00 & $-0.150$ & 14 &   1& syn(166.1) & $ 0.60$ \\
\ion{La}{II} & 3713.54 &  0.17 & $-0.800$ & 15 &   2& syn(12.2) & $-0.48$ \\
\ion{La}{II} & 3794.77 &  0.24 & $ 0.210$ & 15 &   2& syn(45.5) & $-0.53$ \\
\ion{La}{II} & 3849.01 &  0.00 & $-0.450$ & 15 &   2& syn(32.4) & $-0.47$ \\
\ion{La}{II} & 3949.10 &  0.40 & $ 0.490$ & 15 &   2&    syn    & $-0.54$ \\
\ion{La}{II} & 3988.51 &  0.40 & $ 0.080$ & 15 &   2&    syn    & $-0.44$ \\
\ion{La}{II} & 3995.75 &  0.17 & $-0.060$ & 15 &   2&    syn    & $-0.44$ \\
\ion{La}{II} & 4086.71 &  0.00 & $-0.070$ & 15 &   2& syn(52.7) & $-0.33$ \\
\ion{La}{II} & 4123.22 &  0.32 & $ 0.130$ & 15 &   2& syn(47.7) & $-0.44$ \\
\ion{La}{II} & 4322.50 &  0.17 & $-0.930$ & 15 &   2& syn(15.2) & $-0.32$ \\
\ion{La}{II} & 4333.75 &  0.17 & $-0.060$ & 15 &   2&    syn    & $-0.34$ \\
\ion{La}{II} & 4526.11 &  0.77 & $-0.590$ & 15 &   2&    5.9    & $-0.46$ \\
\ion{La}{II} & 4920.98 &  0.13 & $-3.375$ & 15 &   2& syn(31.3) & $-0.13$ \\
\ion{La}{II} & 4921.77 &  0.24 & $-1.139$ & 15 &   2& syn(30.9) & $-0.08$ \\
\ion{Ce}{II} & 3539.08 &  0.32 & $-0.380$ & 16 &   &    8.5    & $-0.10$ \\
\ion{Ce}{II} & 3577.46 &  0.47 & $ 0.210$ & 16 &   &   14.2    & $-0.26$ \\
\ion{Ce}{II} & 3655.84 &  0.32 & $-0.020$ & 16 &   &   14.6    & $-0.23$ \\
\ion{Ce}{II} & 3940.33 &  0.32 & $-0.270$ & 16 &   &    7.9    & $-0.33$ \\
\ion{Ce}{II} & 3942.15 &  0.00 & $-0.220$ & 16 &   &   26.6    & $-0.09$ \\
\ion{Ce}{II} & 3942.74 &  0.86 & $ 0.730$ & 16 &   &   17.8    & $-0.30$ \\
\ion{Ce}{II} & 3960.91 &  0.32 & $-0.400$ & 16 &   &   10.6    & $-0.06$ \\
\ion{Ce}{II} & 3964.50 &  0.32 & $-0.540$ & 16 &   &    syn    & $-0.17$ \\
\ion{Ce}{II} & 3984.67 &  0.96 & $ 0.110$ & 16 &   &    7.9    & $ 0.01$ \\
\ion{Ce}{II} & 3992.38 &  0.45 & $-0.170$ & 16 &   &    9.1    & $-0.23$ \\
\ion{Ce}{II} & 3999.24 &  0.09 & $ 0.232$ & 16 &   &   24.1    & $-0.51$ \\
\ion{Ce}{II} & 4003.77 &  0.93 & $ 0.300$ & 16 &   &   10.5    & $-0.07$ \\
\ion{Ce}{II} & 4014.90 &  0.53 & $ 0.140$ & 16 &   &    8.5    & $-0.48$ \\
\ion{Ce}{II} & 4031.33 &  0.32 & $-0.080$ & 16 &   &   12.9    & $-0.29$ \\
\ion{Ce}{II} & 4053.50 &  0.00 & $-0.710$ & 16 &   &   13.4    & $-0.01$ \\
\ion{Ce}{II} & 4073.47 &  0.48 & $ 0.230$ & 16 &   &   20.3    & $-0.18$ \\
\ion{Ce}{II} & 4083.22 &  0.70 & $ 0.270$ & 16 &   &   14.1    & $-0.16$ \\
\ion{Ce}{II} & 4115.37 &  0.92 & $ 0.100$ & 16 &   &    5.9    & $-0.17$ \\
\ion{Ce}{II} & 4118.14 &  0.70 & $ 0.019$ & 16 &   &   11.6    & $-0.02$ \\
\ion{Ce}{II} & 4120.83 &  0.32 & $-0.210$ & 16 &   &   11.6    & $-0.23$ \\
\ion{Ce}{II} & 4127.36 &  0.68 & $ 0.350$ & 16 &   &   15.0    & $-0.24$ \\
\ion{Ce}{II} & 4137.65 &  0.52 & $ 0.440$ & 16 &   &   26.0    & $-0.20$ \\
\ion{Ce}{II} & 4145.00 &  0.70 & $ 0.130$ & 16 &   &    8.0    & $-0.32$ \\
\ion{Ce}{II} & 4165.60 &  0.91 & $ 0.530$ & 16 &   &   14.7    & $-0.17$ \\
\ion{Ce}{II} & 4222.60 &  0.12 & $ 0.020$ & 16 &   &   24.2    & $-0.29$ \\
\ion{Ce}{II} & 4418.78 &  0.86 & $ 0.280$ & 16 &   &   10.8    & $-0.16$ \\
\ion{Ce}{II} & 4486.91 &  0.30 & $-0.260$ & 16 &   &   17.6    & $-0.02$ \\
\ion{Ce}{II} & 4539.74 &  0.33 & $-0.020$ & 16 &   &   17.8    & $-0.23$ \\
\ion{Ce}{II} & 4562.36 &  0.48 & $ 0.230$ & 16 &   &   25.0    & $-0.11$ \\
\ion{Ce}{II} & 4572.28 &  0.68 & $-0.290$ & 16 &   &   16.4    & $ 0.41$ \\
\ion{Ce}{II} & 4593.93 &  0.70 & $ 0.110$ & 16 &   &   13.3    & $-0.09$ \\
\ion{Ce}{II} & 4628.16 &  0.52 & $ 0.200$ & 16 &   &   22.7    & $-0.10$ \\
\ion{Pr}{II} & 3964.81 &  0.05 & $ 0.121$ & 17  &   3&    syn    & $-0.72$ \\
\ion{Pr}{II} & 3965.25 &  0.20 & $ 0.135$ & 17 &   3&    syn    & $-0.62$ \\
\ion{Pr}{II} & 4179.32 &  0.20 & $ 0.480$ & 17 &   3&    syn    & $-0.65$ \\
\ion{Pr}{II} & 4189.49 &  0.37 & $ 0.380$ & 17 &   3&    syn    & $-0.67$ \\
\ion{Pr}{II} & 4222.95 &  0.05 & $ 0.270$ & 17 &   3&    syn    & $-0.65$ \\
\ion{Pr}{II} & 4408.81 &  0.00 & $ 0.180$ & 17 &   3&    syn    & $-0.72$ \\
\ion{Nd}{II} & 3780.38 &  0.47 & $-0.350$ & 18 &   &   12.1    & $-0.06$ \\
\ion{Nd}{II} & 3784.24 &  0.38 & $ 0.150$ & 18 &   &   31.7    & $-0.09$ \\
\ion{Nd}{II} & 3784.84 &  0.06 & $-1.040$ & 18 &   &     8.0    & $-0.04$ \\
\ion{Nd}{II} & 3826.41 &  0.06 & $-0.410$ & 18 &   &   23.3    & $-0.11$ \\
\ion{Nd}{II} & 3838.98 &  0.00 & $-0.240$ & 18 &   &   34.6    & $-0.08$ \\
\ion{Nd}{II} & 3900.22 &  0.47 & $ 0.100$ & 18 &   &   22.8    & $-0.17$ \\
\ion{Nd}{II} & 3973.26 &  0.63 & $ 0.360$ & 18 &   &   25.2    & $-0.20$ \\
\ion{Nd}{II} & 3990.10 &  0.47 & $ 0.130$ & 18 &   &   33.2    & $ 0.04$ \\
\ion{Nd}{II} & 4004.01 &  0.06 & $-0.736$ & 18 &   &   23.3    & $ 0.19$ \\
\ion{Nd}{II} & 4013.22 &  0.18 & $-1.100$ & 18 &   &    4.5    & $-0.14$ \\
\ion{Nd}{II} & 4018.82 &  0.06 & $-0.850$ & 18 &   &   13.6    & $ 0.00$ \\
\ion{Nd}{II} & 4021.33 &  0.32 & $-0.100$ & 18 &   &   26.7    & $-0.06$ \\
\ion{Nd}{II} & 4022.98 &  0.20 & $-1.400$ & 18 &   &   24.9    & $ 1.06$ \\
\ion{Nd}{II} & 4041.06 &  0.47 & $-0.530$ & 18 &   &    8.6    & $-0.08$ \\
\ion{Nd}{II} & 4043.59 &  0.32 & $-0.710$ & 18 &   &    7.4    & $-0.14$ \\
\ion{Nd}{II} & 4051.14 &  0.38 & $-0.300$ & 18 &   &   20.1    & $ 0.03$ \\
\ion{Nd}{II} & 4059.95 &  0.20 & $-0.520$ & 18 &   &   18.1    & $-0.01$ \\
\ion{Nd}{II} & 4061.08 &  0.47 & $ 0.550$ & 18 &   &   53.1    & $ 0.06$ \\
\ion{Nd}{II} & 4069.26 &  0.06 & $-0.570$ & 18 &   &   24.3    & $ 0.04$ \\
\ion{Nd}{II} & 4109.45 &  0.32 & $ 0.350$ & 18 &   &   60.7    & $ 0.27$ \\
\ion{Nd}{II} & 4133.35 &  0.32 & $-0.490$ & 18 &   &   14.3    & $-0.05$ \\
\ion{Nd}{II} & 4211.29 &  0.20 & $-0.860$ & 18 &   &   10.0    & $-0.00$ \\
\ion{Nd}{II} & 4214.60 &  0.18 & $-1.180$ & 18 &   &    4.2    & $-0.12$ \\
\ion{Nd}{II} & 4232.37 &  0.06 & $-0.470$ & 18 &   &   25.6    & $-0.05$ \\
\ion{Nd}{II} & 4358.16 &  0.32 & $-0.160$ & 18 &   &   30.9    & $ 0.05$ \\
\ion{Nd}{II} & 4446.38 &  0.20 & $-0.350$ & 18 &   &   26.6    & $-0.00$ \\
\ion{Nd}{II} & 4462.98 &  0.56 & $ 0.040$ & 18 &   &   28.2    & $ 0.05$ \\
\ion{Nd}{II} & 4542.60 &  0.74 & $-0.280$ & 18 &   &    9.3    & $-0.04$ \\
\ion{Nd}{II} & 4567.60 &  0.20 & $-1.250$ & 18 &   &    2.9    & $-0.23$ \\
\ion{Nd}{II} & 4579.31 &  0.74 & $-0.480$ & 18 &   &    7.2    & $ 0.04$ \\
\ion{Nd}{II} & 4597.01 &  0.20 & $-1.150$ & 18 &   &    8.1    & $ 0.14$ \\
\ion{Nd}{II} & 4645.76 &  0.56 & $-0.760$ & 18 &   &    7.9    & $ 0.14$ \\
\ion{Nd}{II} & 4706.54 &  0.00 & $-0.710$ & 18 &   &   23.7    & $ 0.02$ \\
\ion{Nd}{II} & 4715.59 &  0.20 & $-0.900$ & 18 &   &    9.5    & $-0.04$ \\
\ion{Nd}{II} & 4859.03 &  0.32 & $-0.440$ & 18 &   &   20.1    & $ 0.01$ \\
\ion{Sm}{II} & 3706.75 &  0.49 & $-0.600$ & 19 &   &    8.2    & $-0.24$ \\
\ion{Sm}{II} & 3760.71 &  0.19 & $-0.400$ & 19 &   &   23.7    & $-0.23$ \\
\ion{Sm}{II} & 3793.98 &  0.10 & $-0.670$ & 19 &   &   15.9    & $-0.29$ \\
\ion{Sm}{II} & 3896.97 &  0.04 & $-0.670$ & 19 &   &   15.5    & $-0.39$ \\
\ion{Sm}{II} & 3993.31 &  0.04 & $-0.930$ & 19 &   &   13.7    & $-0.21$ \\
\ion{Sm}{II} & 4023.22 &  0.04 & $-0.930$ & 19 &   &   11.3    & $-0.31$ \\
\ion{Sm}{II} & 4206.12 &  0.38 & $-0.720$ & 19 &   &    6.8    & $-0.40$ \\
\ion{Sm}{II} & 4318.93 &  0.28 & $-0.250$ & 19 &   &   23.0    & $-0.36$ \\
\ion{Sm}{II} & 4420.52 &  0.33 & $-0.430$ & 19 &   &   30.2    & $ 0.043$ \\
\ion{Sm}{II} & 4499.48 &  0.25 & $-0.870$ & 19 &   &    9.9    & $-0.25$ \\
\ion{Sm}{II} & 4519.63 &  0.54 & $-0.350$ & 19 &    &  12.0    & $-0.34$ \\
\ion{Sm}{II} & 4523.91 &  0.43 & $-0.390$ & 19 &   &   14.0    & $-0.34$ \\
\ion{Sm}{II} & 4537.94 &  0.48 & $-0.480$ & 19 &   &   10.6    & $-0.34$ \\
\ion{Sm}{II} & 4577.69 &  0.25 & $-0.650$ & 19 &    &  13.7    & $-0.31$ \\
\ion{Sm}{II} & 4591.81 &  0.19 & $-1.120$ & 19 &   &    5.1    & $-0.40$ \\
\ion{Sm}{II} & 4642.23 &  0.38 & $-0.460$ & 19 &   &   16.8    & $-0.25$ \\
\ion{Sm}{II} & 4815.81 &  0.19 & $-0.820$ & 19 &   &    9.6    & $-0.42$ \\
\ion{Eu}{II} & 3724.93 &  0.00 & $-0.090$ & 20 & 4 & syn(98.6) & $-0.65$ \\
\ion{Eu}{II} & 3819.67 &  0.00 & $ 0.510$ & 20 & 4 &    syn    & $-0.65$ \\
\ion{Eu}{II} & 3907.11 &  0.21 & $ 0.170$ & 20 & 4 & syn(75.7) & $-0.65$:\\
\ion{Eu}{II} & 3930.50 &  0.21 & $ 0.270$ & 20 & 4 & syn(91.1) & $-0.68$ \\
\ion{Eu}{II} & 3971.97 &  0.21 & $ 0.270$ & 20 & 4 &    syn    & $-0.77$ \\
\ion{Eu}{II} & 4129.73 &  0.00 & $ 0.220$ & 20 & 4 & syn(133.6) & $-0.65$ \\
\ion{Eu}{II} & 4205.04 &  0.00 & $ 0.210$ & 20 & 4 &    syn    & $-0.65$ \\
\ion{Eu}{II} & 4435.58 &  0.21 & $-0.110$ & 20 & 4 &    0.0    & $ 0.00$ \\
\ion{Eu}{II} & 4522.58 &  0.21 & $-0.670$ & 20 & 4 &    0.0    & $ 0.00$ \\
\ion{Gd}{II} & 3331.39 &  0.00 & $-0.140$ & 1 &   &   25.9    & $-0.37$ \\
\ion{Gd}{II} & 3424.59 &  0.35 & $-0.170$ & 1 &   &   11.6    & $-0.42$ \\
\ion{Gd}{II} & 3439.21 &  0.38 & $ 0.150$ & 1 &   &   26.5    & $-0.23$ \\
\ion{Gd}{II} & 3451.24 &  0.38 & $-0.050$ & 1 &   &   11.5    & $-0.52$ \\
\ion{Gd}{II} & 3454.91 &  0.03 & $-0.480$ & 1 &    &  16.1    & $-0.31$ \\
\ion{Gd}{II} & 3481.80 &  0.49 & $ 0.230$ & 1 &   &   19.6    & $-0.38$ \\
\ion{Gd}{II} & 3549.36 &  0.24 & $ 0.260$ & 1 &   &   30.0    & $-0.43$ \\
\ion{Gd}{II} & 3557.06 &  0.60 & $ 0.210$ & 1 &   &   15.1    & $-0.40$ \\
\ion{Gd}{II} & 3697.73 &  0.03 & $-0.280$ & 1 &   &    syn    & $-0.24$ \\
\ion{Gd}{II} & 3712.70 &  0.38 & $ 0.150$ & 1 &   &    syn    & $-0.29$ \\
\ion{Gd}{II} & 3768.40 &  0.08 & $ 0.360$ & 1 &   &   47.8    & $-0.37$ \\
\ion{Gd}{II} & 3796.38 &  0.03 & $ 0.140$ & 1 &   &   49.4    & $-0.17$ \\
\ion{Gd}{II} & 3844.58 &  0.14 & $-0.400$ & 1 &   &   23.4    & $-0.14$ \\
\ion{Gd}{II} & 3916.51 &  0.60 & $ 0.060$ &  2 &   &   16.3    & $-0.29$ \\
\ion{Gd}{II} & 3973.98 &  0.60 & $-0.400$ & 1 &   &    4.7    & $-0.45$ \\
\ion{Gd}{II} & 4037.32 &  0.66 & $-0.020$ & 1 &    &  16.3    & $-0.16$ \\
\ion{Gd}{II} & 4037.89 &  0.56 & $-0.230$ & 1 &    &   7.4    & $-0.46$ \\
\ion{Gd}{II} & 4085.56 &  0.73 & $ 0.070$ & 1 &    &   9.3    & $-0.46$ \\
\ion{Gd}{II} & 4191.07 &  0.43 & $-0.570$ & 1 &   &   10.0    & $-0.15$ \\
\ion{Gd}{II} & 4215.02 &  0.43 & $-0.580$ & 1 &   &   10.6    & $-0.11$ \\
\ion{Tb}{II} & 3472.80 &  0.13 & $-0.100$ & 21 & 5 &    0.0    & $ 0.00$ \\
\ion{Tb}{II} & 3568.45 &  0.00 & $ 0.360$ & 21 & 5 &    syn    & $-1.06$ \\
\ion{Tb}{II} & 3600.38 &  0.64 & $ 0.600$ & 21 & 5 &    syn    & $-1.14$ \\
\ion{Tb}{II} & 3658.89 &  0.13 & $-0.010$ & 21 & 5 & syn(7.5)  & $-1.01$ \\
\ion{Tb}{II} & 3702.85 &  0.13 & $ 0.440$ & 21 & 5 & syn(25.6) & $-1.04$ \\
\ion{Tb}{II} & 3848.73 &  0.00 & $ 0.280$ & 21 & 5 &    syn    & $-1.04$ \\
\ion{Tb}{II} & 3874.17 &  0.00 & $ 0.270$ & 21 & 5 &    syn    & $-0.96$ \\
\ion{Tb}{II} & 3899.19 &  0.37 & $ 0.330$ & 21 & 5 &    syn    & $-1.04$ \\
\ion{Tb}{II} & 4002.57 &  0.64 & $ 0.100$ & 21 & 5 &    syn    & $-1.01$:\\
\ion{Tb}{II} & 4005.47 &  0.13 & $-0.020$ & 21 & 5 & syn(12.3) & $-1.11$ \\
\ion{Dy}{II} & 3407.80 &  0.00 & $ 0.180$ & 1 &   &   56.0    & $-0.06$ \\
\ion{Dy}{II} & 3434.37 &  0.00 & $-0.450$ & 1 &   &   32.8    & $-0.10$ \\
\ion{Dy}{II} & 3445.57 &  0.00 & $-0.150$ & 1 &   &   44.9    & $-0.08$ \\
\ion{Dy}{II} & 3454.32 &  0.10 & $-0.140$ & 1 &   &   36.7    & $-0.20$ \\
\ion{Dy}{II} & 3460.97 &  0.00 & $-0.070$ & 1 &   &   46.8    & $-0.12$ \\
\ion{Dy}{II} & 3536.02 &  0.54 & $ 0.530$ & 1 &   &   46.0    & $-0.14$ \\
\ion{Dy}{II} & 3538.52 &  0.00 & $-0.020$ & 1 &   &   54.7    & $ 0.05$ \\
\ion{Dy}{II} & 3546.83 &  0.10 & $-0.550$ & 1 &   &   26.5    & $-0.06$ \\
\ion{Dy}{II} & 3550.22 &  0.59 & $ 0.270$ & 1 &   &   30.3    & $-0.22$ \\
\ion{Dy}{II} & 3559.30 &  1.22 & $-0.280$ & 1 &   &    3.5    & $-0.11$ \\
\ion{Dy}{II} & 3563.15 &  0.10 & $-0.360$ & 1 &   &  34.4    & $-0.06$ \\
\ion{Dy}{II} & 3694.81 &  0.10 & $-0.110$ & 1 &   &   47.2    & $-0.06$ \\
\ion{Dy}{II} & 3708.22 &  0.59 & $-0.880$ & 1 &   &    5.9    & $-0.04$ \\
\ion{Dy}{II} & 3747.82 &  0.10 & $-0.810$ & 1 &   &   15.9    & $-0.18$ \\
\ion{Dy}{II} & 3788.44 &  0.10 & $-0.570$ & 1 &   &   29.4    & $-0.05$ \\
\ion{Dy}{II} & 3869.86 &  0.00 & $-0.940$ & 2 &   &   17.4    & $-0.14$ \\
\ion{Dy}{II} & 3983.65 &  0.54 & $-0.310$ & 1 &    &  21.6    & $-0.04$ \\
\ion{Dy}{II} & 3996.69 &  0.59 & $-0.260$ & 1 &   &   19.3    & $-0.10$ \\
\ion{Dy}{II} & 4073.12 &  0.54 & $-0.320$ & 1 &   &   23.2    & $ 0.00$ \\
\ion{Dy}{II} & 4077.97 &  0.10 & $-0.040$ & 1 &   &   63.8    & $ 0.22$ \\
\ion{Dy}{II} & 4103.31 &  0.10 & $-0.380$ & 1 &   &   44.5    & $ 0.05$ \\
\ion{Dy}{II} & 4468.14 &  0.10 & $-1.670$ & 1 &   &    4.5    & $-0.03$ \\
\ion{Ho}{II} & 3796.75 &  0.00 & $ 0.160$ & 22 & 6 &   syn    & $-0.78$ \\
\ion{Ho}{II} & 3810.74 &  0.00 & $ 0.190$ & 22 & 6 &    syn    & $-0.77$ \\
\ion{Ho}{II} & 4045.47 &  0.00 & $-0.050$ & 22 & 6 &    syn    & $-0.70$ \\
\ion{Ho}{II} & 4152.59 &  0.08 & $-0.930$ & 22 & 6 &    0.0    & $ 0.00$ \\
\ion{Er}{II} & 3332.70 &  0.89 & $ 0.070$ & 23 &   &  11.0    & $-0.37$ \\
\ion{Er}{II} & 3499.10 &  0.05 & $ 0.290$ & 23 &   &   48.0    & $-0.49$ \\
\ion{Er}{II} & 3559.89 &  0.00 & $-0.690$ & 23 &   &   22.6    & $-0.25$ \\
\ion{Er}{II} & 3692.65 &  0.05 & $ 0.280$ & 23 &   &   59.2    & $-0.28$ \\
\ion{Er}{II} & 3729.52 &  0.00 & $-0.590$ & 23 &   &   28.4    & $-0.27$ \\
\ion{Er}{II} & 3786.84 &  0.00 & $-0.520$ & 23 &   &   35.1    & $-0.19$ \\
\ion{Er}{II} & 3830.48 &  0.00 & $-0.220$ & 23 &   &   45.1    & $-0.26$ \\
\ion{Er}{II} & 3896.23 &  0.05 & $-0.120$ & 23 &   &   53.4    & $-0.11$ \\
\ion{Er}{II} & 3938.63 &  0.00 & $-0.520$ & 23 &   &   27.5    & $-0.39$ \\
\ion{Tm}{II} & 3462.20 &  0.00 & $ 0.030$ & 1 &   & syn(24.0) & $-1.24$ \\
\ion{Tm}{II} & 3700.26 &  0.03 & $-0.290$ & 1 &   & syn(16.7) & $-1.18$ \\
\ion{Tm}{II} & 3701.36 &  0.00 & $-0.540$ & 1 &   & syn(10.8) & $-1.20$ \\
\ion{Tm}{II} & 3761.33 &  0.00 & $-0.430$ & 1 &   &    syn    & $-1.25$ \\
\ion{Tm}{II} & 3795.76 &  0.03 & $-0.230$ & 1 &   & syn(16.3) & $-1.15$ \\
\ion{Tm}{II} & 3848.02 &  0.00 & $-0.130$ & 1 &   & syn(27.1) & $-1.10$ \\
\ion{Yb}{II} & 3289.37 &  0.00 & $ 0.020$ & 1 & 7 &    syn    & $-0.20$ \\
\ion{Yb}{II} & 3694.19 &  0.00 & $-0.300$ & 1 & 7 &    syn    & $-0.30$ \\
\ion{Lu}{II} & 3397.02 &  1.46 & $-0.190$ & 1 & 1  &    syn    & $-0.72$ \\
\ion{Lu}{II} & 3472.45 &  1.54 & $-0.190$ & 1 & 1  &    syn    & $-1.02$:\\
\ion{Lu}{II} & 3554.39 &  2.15 & $ 0.140$ &  1 & 1  &    syn    & $-1.02$ \\
%\ion{Hf}{II} & 3399.79 &  0.00 & $-0.490$ & & syn(34.6) & $-0.63$ \\ old loggf
%\ion{Hf}{II} & 3719.28 &  0.61 & $-0.870$ & &    syn    & $-0.57$ \\  old loggf
%\ion{Hf}{II} & 3793.38 &  0.38 & $-0.950$ & &    syn    & $-0.60$ \\ old loggf
\ion{Hf}{II} &3389.829  &  0.45 & $-0.780$ & 24 &   &    syn    & $-0.75$ \\    
\ion{Hf}{II} &3399.793  &  0.00 & $-0.570$ & 24 &   &syn(34.6)& $-0.62$ \\    
\ion{Hf}{II} &3479.289  &  0.38 & $-1.000$ & 24 &   &    syn    & $-0.60$ \\   
\ion{Hf}{II} &3505.219  &  1.04 & $-0.140$ & 24 &   &    syn    & $-0.75$ \\    
\ion{Hf}{II} &3561.659  &  0.00 & $-0.870$ & 24 &   &    syn    & $-0.80$ \\  
\ion{Hf}{II} &3719.276  &  0.61 & $-0.810$ & 24 &   &    syn    & $-0.65$ \\    
\ion{Hf}{II} &3793.379  &  0.38 & $-1.110$ & 24 &   &    syn    & $-0.50$ \\    
\ion{Hf}{II} &3918.094  &  0.45 & $-1.140$ & 24 &   &    syn    & $-0.60$ \\    
\ion{Hf}{II} &4093.155  &  0.45 & $-1.150$ & 24 &   &    syn    & $-0.65$ \\    
\ion{Os}{I } & 3267.95 &  0.00 & $-1.080$ & 25 &    &    syn(13.0) & $ 0.36$ \\
\ion{Os}{I } & 3301.57 &  0.00 & $-0.743$ & 25 &     &    syn    & $ 0.27$ \\
\ion{Os}{I } & 3528.60 &  0.00 & $-1.740$ & 25 &     & syn(7.5)  & $ 0.47$ \\
\ion{Os}{I } & 4135.77 &  0.52 & $-1.260$ & 25 &     & syn(6.2)  & $ 0.62$ \\
\ion{Os}{I } & 4260.85 &  0.00 & $-1.440$ & 25 &     & syn(8.2)  & $ 0.37$ \\
\ion{Ir}{I } & 3513.65 &  0.00 & $-1.246$ & 25 &     & syn(25.7) & $ 0.25$ \\
\ion{Ir}{I } & 3800.12 &  0.00 & $-1.489$ & 25 &     & syn(15.2) & $ 0.22$ \\
\ion{Pt}{I } & 3301.86 &  0.81 & $-0.770$ & 1 &     &    syn    & $ 0.10$:\\
\ion{Pt}{I } & 3315.04 &  0.00 & $-2.580$ & 1 &     &    syn    & $ 0.78$:\\
\ion{Pb}{I } & 4057.81 &  1.32 & $-0.220$ & 26 &     &    syn    & $ 0.05$:\\
\ion{Th}{II} & 3351.23 &  0.19 & $-0.600$ & 27 &     & syn(10.7) & $-0.90$ \\
\ion{Th}{II} & 3435.98 &  0.00 & $-0.670$ & 27 &     & syn(4.7)  & $-1.05$ \\
\ion{Th}{II} & 3469.92 &  0.51 & $-0.129$ & 27 &     &    syn    & $-1.05$ \\
\ion{Th}{II} & 3675.57 &  0.19 & $-0.840$ & 27 &     & syn(2.2)  & $-1.00$ \\
\ion{Th}{II} & 4019.13 &  0.00 & $-0.228$ & 27 &     & syn(16.0) & $-1.25$ \\
\ion{Th}{II} & 4086.52 &  0.00 & $-0.929$ & 27 &     &    syn    & $-1.15$ \\
\ion{ U}{II} & 3859.57 &  0.04 & $-0.067$ & 28 &     &    syn    & $-2.20$:\\
\hline 
\end{longtable} 
%}% End \longtab 

\end{appendix}

\end{document}